\documentclass[A4paper]{article}

\usepackage[T1]{fontenc}
\usepackage{textcomp}
\usepackage[utf8]{inputenc}
\usepackage{color}
\usepackage{amsmath}
\usepackage{amsthm}
\usepackage{amssymb}
\usepackage{graphicx}
\usepackage{esint}

\makeatletter

\pdfpageheight\paperheight
\pdfpagewidth\paperwidth

\newcommand*\LyXZeroWidthSpace{\hspace{0pt}}
\providecommand{\tabularnewline}{\\}

\numberwithin{equation}{section}
\numberwithin{figure}{section}
\theoremstyle{plain}
\newtheorem*{prop*}{\protect\propositionname}
\theoremstyle{plain}
\newtheorem*{cor*}{\protect\corollaryname}

\@ifundefined{date}{}{\date{}}

\usepackage{color}
\usepackage{amsfonts}\usepackage{amsthm}\usepackage{amscd}\numberwithin{equation}{section}
\usepackage{bm}
\usepackage[all]{xy}

\usepackage{algorithm}
\usepackage{algorithmicx}
\usepackage{algpseudocode}
\usepackage[most]{tcolorbox}
\usepackage{float}
\usepackage{lipsum}  
\tcbset{
  highlight algorithm/.style={
    colback=gray!10,
    colframe=black,
    boxrule=0.5pt,
    arc=2pt,
    left=6pt,
    right=6pt,
    top=6pt,
    bottom=6pt,
    width=\textwidth,
    sharp corners,
  }
}
\usepackage{parskip} 

\makeatother

\usepackage[english]{babel}
\providecommand{\corollaryname}{Corollary}
\providecommand{\propositionname}{Proposition}

\begin{document}
\begin{flushleft}
\textbf{OUJ-FTC-14}\\
 \textbf{OCHA-PP-379}\\
 
\par\end{flushleft}

\thispagestyle{empty} \vspace*{-15mm}

\vspace{15mm}

\begin{center}
{\Large\textbf{Inflation Model Based on Virasoro Squeezing}}\textbf{ }
\par\end{center}

\vspace{10pt}

\begin{center}
Kenji Ebata\textsuperscript{{*}}, So Katagiri\textsuperscript{{*}}\footnote{So.Katagiri@gmail.com},
Yoshiki Matsuoka$^{*}$, Takaaki Sehara$^{*}$, and Akio Sugamoto$^{\dagger}$
\par\end{center}

\begin{center}
\textit{$^{*}$Nature and Environment, Faculty of Liberal Arts, The
Open University of Japan, Chiba 261-8586, Japan} \\
\par\end{center}

\begin{center}
\textit{$^{\dagger}$Department of Physics, Graduate School of Humanities
and Sciences, Ochanomizu University, 2-1-1 Otsuka, Bunkyo-ku, Tokyo
112-8610, Japan }
\par\end{center}

\begin{center}
\begin{minipage}[c]{14cm}%
\begin{abstract}
\setlength{\parindent}{0pt}
\setlength{\parskip}{1em}

We propose a novel mechanism for realizing slow-roll inflation that
is fully consistent with observational data, based on conformal transformations
acting exclusively on a complex scalar field---without coupling to
the gravitational sector. These transformations generically produce
a plateau in the inflaton potential\color{black}, as suggested by the asymptotic behavior of holomorphic transformations and the maximum modulus theorem\color{black},
 thereby naturally satisfying the slow-roll conditions.

Our framework utilizes squeezing operations generated by the Virasoro
algebra without central extension, as developed in our earlier work.
The resulting inflationary potentials depend on the Virasoro mode
$n$, the power $m$ of the original potential, and the squeezing
parameter $\theta$.

We present approximate analytical expressions at leading order for
the special case $n=-2$, and perform numerical analyses for both
$n=-2$ and other values of $n$. These reveal parameter regimes in
which the predicted cosmological observables $(n_{s},r)$ align remarkably
well with current CMB measurements.
\end{abstract}
\end{minipage}
\par\end{center}

\section{Introduction}

The question of the origin of the universe has been a subject of profound
human curiosity since antiquity. While this inquiry once belonged
to the realm of myth, the discovery of the cosmic microwave background
(CMB) has, for the first time, rendered the origin of the universe
amenable to rigorous scientific investigation.

Modern standard cosmology is based on the theory of inflation, which
posits a period of rapid accelerated expansion in the early universe\footnote{See \cite{PLACNK2018}. See also\cite{Ade2021}. For the theoretical
background, refer to \cite{Hu2002}.}. This paradigm provides a natural explanation for key problems that
plagued previous models, including the observed spatial flatness of
the universe and the strong correlations among galaxies that reside
in causally disconnected regions.

Furthermore, a key prediction of inflation---the generation of scalar
and tensor perturbations that manifest as temperature anisotropies
in the CMB---has been confirmed by high-precision observations, thereby
greatly enhancing the credibility of the inflationary paradigm. The
increasing precision of CMB measurements has now enabled a wide range
of inflationary models to be quantitatively tested and discriminated\cite{PLACNK2018, Ade2021}.

Among the well-known scenarios are the Starobinsky model based on
$R^{2}$ gravity\cite{Starobinsky1980} and Linde's chaotic inflation
model\cite{Linde1983}. These models have mutually influenced each
other, giving rise to a wide variety of inflationary scenarios, including
Higgs inflation\cite{Kehagias2014}, $\alpha$-attractors\cite{Kallosh2022}
and string-inspired models\cite{Silverstein2008,Shamit_Kachru_2003}.
Recent reviews provide comprehensive summaries of these developments\cite{Linde1983, Kallosh2025}.

In particular, the $\alpha$-attractor models proposed by Kallosh
and Linde\cite{Kallosh2013} can be understood as extensions of the
Starobinsky scenario. By constructing actions in which the gravitational
and scalar sectors are coupled in a conformally invariant manner,
these models can generate a family of inflationary potentials consistent
with observational data. Theoretical advances and observational constraints
on $\alpha$-attractors have been intensively discussed in recent
literature\cite{Kallosh2022,Kubota2023,Bhattacharya2023,Laura2023CMB}.

A common feature among these models is that the inflaton potential
enabling slow-roll inflation is typically constructed by design. In
other words, the realization of slow-roll conditions is essentially
a consequence of the deliberate choice of potential form, rather than
being dictated by any underlying physical necessity.

In this work, we propose a fundamentally different approach to this
issue by introducing a new framework in which slow-roll inflation
emerges naturally through the application of specific modes of conformal
transformations to the scalar field alone. Notably, our construction
does not require any special coupling to gravity, setting it apart
from conventional models\footnote{The scalar field in our model is not conformally coupled to gravity; only the scalar sector undergoes Virasoro squeezing, while the metric remains fixed.}.

This conformal transformation is realized via a higher-order squeezing
operation, termed Virasoro squeezing\cite{Katagiri2019}\footnote{For subsequent developments of Virasoro squeezing and its algebraic
generalizations, see \cite{Katagir2023,NietoChaupis2025}.}, which is generated by the Virasoro algebra $L_{n}$ without central
extension, as developed in our previous work. Depending on the Virasoro
mode $n$, the original potential $V(\phi)$, and the squeezing strength
$\theta$, the transformed potential typically develops a plateau
structure in the asymptotic regime, as suggested by the asymptotic behavior of holomorphic transformations and the maximum modulus theorem in complex analysis. This plateau naturally facilitates slow-roll
inflation for suitable choices of $n$ and $\theta$. 

In particular, we provide an approximate analytical expression for
the special case of $n=-2$, valid to leading order. This serves as
a useful benchmark, but higher-order corrections must be treated numerically.
\color{black}
The $n=-2$ case should be regarded as an analytic benchmark, since it reproduces the Starobinsky / $\alpha$-attractor t-model only at leading order.
\color{black}

\color{black}
The novelty of the present framework is twofold. First, plateau-like potentials are generated by Virasoro squeezing of the scalar sector, without introducing a specially designed non-minimal coupling to gravity. Second, the Virasoro mode n provides a systematic way to generate a family of inflationary models with mode-dependent predictions for $(n_s,r)$\footnote{The novelty of our approach lies in the fact that, by postulating
Virasoro squeezing of a complex scalar field in the early universe,
the emergence of a slow-roll potential becomes a natural consequence
of the underlying framework. While possible physical mechanisms responsible
for such squeezing include vacuum condensation of the complex scalar
field or geometric friction effects arising from intersecting branes,
a detailed exploration of these origins is beyond the scope of this
work. Accordingly, the present study focuses on whether appropriate
Virasoro squeezing can reproduce the observed CMB fluctuations.}.
\color{black}

The remainder of this paper is organized as follows. In Section 2,
we present the formalism of inflationary models based on Virasoro
squeezing. Section 3 focuses on the special case of $n=-2$, where
we derive analytical solutions and demonstrate that the predicted
observables cover the full range of current experimental constraints,
showing agreement with the $t$-model of the $\alpha$-attractor scenario.
In Section 4, we perform a numerical analysis for other values of
$n$, identifying parameter regions consistent with observations.
Section 5 provides approximate analytical expressions, derived via
perturbative analysis, that capture the characteristic behavior of
the model. Finally, Section 6 summarizes our conclusions and discusses
prospects for future work.

Appendix A details the numerical algorithms used in our computations,
while Appendix B provides the technical details of the perturbative
analysis, and Appendix C discusses the validity range of the squeezing
parameter~$\theta$ in relation to observational constraints and
regularity conditions.

\section{Inflationary Models Based on Virasoro Squeezing}

In this section, we describe Virasoro squeezing---a higher-order
generalization of the squeezing transformation introduced in our previous
work\cite{Katagiri2019}. We explain how applying Virasoro squeezing
to the potential $V(|\phi|)$ of an existing complex scalar field
naturally leads to inflationary models that realize slow-roll inflation.

\subsection{Virasoro Squeezing}

The concept of squeezing based on the Virasoro algebra was first introduced
in our previous work \cite{Katagiri2019}\footnote{We assume that the Virasoro squeezing occurs prior to the onset of inflation, as motivated by external modulation in the early universe (see Eq. (\ref{Houtside})).}. In this framework, the
$n$-th order squeezing is realized by applying the squeezing operator
$\hat{S}_{n}$ either to a quantum state $|\alpha\rangle$ or to an
operator $\hat{O}$, as follows:

\begin{eqnarray}
 &  & |\alpha'\rangle=\hat{S}_{n}^{-1}|\alpha\rangle,\ \hat{O}'=\hat{S}_{n}\hat{O}\hat{S}_{n}^{-1}.
\end{eqnarray}

Here, $\hat{S}_{n}$ is defined in terms of the $n$-th Virasoro generator\footnote{While the concept of $n$-th order Virasoro squeezing was originally
introduced with integer $n$ in our previous work \cite{Katagiri2019},
the formalism admits analytical continuation to real or complex values.
In the present study, we explicitly adopt real-valued $n$ in Sections
4 and 5 to enable flexible model building and detailed comparison
with observational data.}, $\hat{L}_{n}$ and a complex parameter\footnote{Although the squeezing parameter $\theta$ is generally defined as
a complex number, in this study we restrict it to real values when
constructing the inflationary model and performing numerical calculations.
This restriction is imposed to ensure that the resulting potential
remains real-valued, which is necessary for physical predictions.
Importantly, the essential effects of the Virasoro transformation---such
as the emergence of a plateau in the inflaton potential---are preserved
under this real-valued restriction.} $\theta$, as 
\begin{eqnarray}
\hat{S}_{n}=e^{\theta\hat{L}_{n}}.
\end{eqnarray}
The generator $\hat{L}_{n}$, responsible for the conformal transformation,
is given by
\begin{eqnarray}
\hat{L}_{n}=\phi^{n+1}\frac{\partial}{\partial\phi},
\end{eqnarray}
and satisfies the Virasoro algebra, 
\begin{eqnarray}
[\hat{L}_{m},\hat{L}_{n}]=(m-n)\hat{L}_{m+n},
\end{eqnarray}
where the central extension is neglected.\footnote{The central extension may appear as a commutator anomaly, in the case
when the Virasoro generators are given dynamically, in terms of the
creation and annihilation operators of the system.}

The above construction is equally applicable in a classical context,
including the analysis of cosmic inflation.

Since $\hat{L}_{n}$ is a generator of the following infinitesimal
conformal transformation, 
\begin{eqnarray}
\phi\to\phi+\delta\theta\;\phi^{n+1},
\end{eqnarray}
the finite conformal transformation (with a finite $\theta$) can
be written with $\hat{S}_{n}(\theta)=e^{\theta\hat{L}_{n}}$ as follows:
\begin{eqnarray}
\phi'=F_{n}(\phi;\theta)=\hat{S}_{n}(\theta)\phi\hat{S}_{n}(\theta)^{-1}.\label{finite conformal}
\end{eqnarray}

\color{black}

The classical field configuration is identified with the expectation
value

\begin{equation}
\phi_{c}(x)=\langle\Omega|e^{iHt}\phi(x)e^{-iHt}|\Omega\rangle_{J}
\end{equation}

When Virasoro squeezing acts on the vacuum, the vacuum state $|\Omega\rangle$
is transformed as
\begin{equation}
|\Omega_{n}(\theta)\rangle\equiv S_{n}(\theta)^{-1}|\Omega\rangle
\end{equation}

In the corresponding similarity representation, we take the dual state as
\begin{equation}
\langle \Omega_n(\theta)| \equiv \langle \Omega|S_n(\theta).
\end{equation}
This convention allows the state transformation to be equivalently represented as a similarity transformation acting on operators.

The time evolution of the classical configuration in the squeezed
vacuum state is then given by
\begin{equation}
\phi'_{c}(x)=\langle\Omega_{n}(\theta)|e^{iHt}\phi(x)e^{-iHt}|\Omega_{n}(\theta)\rangle_{J}=\langle\Omega|e^{iH^{(n)}(\theta)t}S_{n}(\theta)\phi(x)S_{n}(\theta)^{-1}e^{-iH^{(n)}(\theta)t}|\Omega\rangle_{J}
\end{equation}

\begin{equation}
=\langle\Omega|e^{iH^{(n)}(\theta)t}F_{n}(\phi;\theta)e^{-iH^{(n)}(\theta)t}|\Omega\rangle_{J}
\end{equation}

where
\begin{equation}
H^{(n)}(\theta)\equiv S_{n}(\theta)HS_{n}(\theta)^{-1}.
\end{equation}

In the following, we denote $\phi_{c}$ simply by $\phi$.

It is important to note that Virasoro squeezing is not a symmetry
transformation of the system in general, but rather represents a change
of the vacuum state.

Here,

\begin{equation}
F_{n}(\phi;\theta)\equiv S_{n}(\theta)\phi(x)S_{n}(\theta)^{-1}
\end{equation}

We denote by $\Pi$ the canonical momentum conjugate to $\phi$, and
assume that the Hamiltonian takes the form

\begin{equation}
H=\int d^{3}x\left(a^{-3}\Pi^{\dagger}\Pi+a^{3}V(\phi)\right)
\end{equation}

which will be used in the subsequent discussion.

Under the squeezing transformation, the canonical momentum transforms
as
\begin{equation}
\Pi'^\dagger
\equiv
S_n(\theta)\Pi^\dagger S_n(\theta)^{-1}
=
F_n'(\phi;\theta)^{-1}\Pi^\dagger
\end{equation}

This transformation preserves the canonical structure, in the sense
that

\begin{equation}
\{\phi',\Pi'^\dagger\}_{\mathrm{pb}}=\{\phi,\Pi^\dagger\}_{\mathrm{pb}}.
\end{equation}

Note also that, in general,
\begin{equation}
[H,S_{n}(\theta)]\neq0
\end{equation}

and therefore this transformation is not a symmetry of the system.

Furthermore, the velocity operator transforms as
\begin{equation}
S_{n}(\theta)\dot{\phi}(x)S_{n}(\theta)^{-1}=a^{-3}S_{n}(\theta)\Pi^\dagger S_{n}(\theta)^{-1}=a^{-3}F'{}_{n}^{-1}(\phi;\theta)\Pi^\dagger=F'{}_{n}^{-1}(\phi;\theta)\dot{\phi}.
\end{equation}
\color{black}

We use this expression in building an inflation model.

We note that under a conformal transformation, a wave of the form
$\phi=e^{-i\omega t}$ is mapped to another wave $\phi^{n+1}=e^{-i(n+1)\omega t}$,
which propagates with a frequency increased by a factor of $n+1$.
In periodic systems such as the harmonic oscillator or atomic systems,
this corresponds to a reduction of the period to $T=\frac{T_{0}}{n+1}$
where $T_{0}$ is the original period. As a result, higher-frequency
modes are introduced, leading to a squeezing phenomenon that drastically
alters the shape of the Husimi distribution in phase space (see Figure
2 in Ref. \cite{Katagiri2019}).

To realize such squeezing in a cosmological setting, we assume that,
just before the onset of inflation, the universe undergoes strong
external modulation, whereby higher oscillation modes are injected
into the inflaton field through a time-dependent Hamiltonian $H_{\text{outside}}$.
This Hamiltonian may be modeled as
\begin{eqnarray} \label{Houtside}
H_{\text{outside}}=\frac{1}{2}\;\Omega(t)\;\phi^{n+2},
\end{eqnarray}
where $\Omega(t)$ represents an external driving variable. For further
details on higher-order squeezing, we refer the reader to Ref. \cite{Katagiri2019}.

\color{black}
As in ordinary squeezing, the microscopic origin of the external modulation depends on the underlying system. In the present work we do not attempt to derive this modulation from a UV-complete theory; instead, we study the inflationary consequences once such a squeezing operation is realized.
\color{black}

\subsection{Virasoro Squeezing Inflation Model}

We consider the modulus of a complex scalar field as the inflaton,
coupled to the spacetime metric $g_{\mu\nu}(x)$ via the action
\begin{eqnarray}
S_{\phi}=\int d^{4}x\sqrt{-g(x)}\left\{ g^{\mu\nu}(x)\partial_{\mu}\phi(x)^{\dagger}\partial_{\nu}\phi(x)-V_{m}(|\phi(x)|^{2})\right\} ,
\end{eqnarray}
where the potential is given by

\begin{equation}
V_{m}(|\phi(x)|^{2})=\frac{\lambda_{m}}{m}|\phi(x)|^{m}.
\end{equation}

For a homogeneous and isotropic universe, the metric is given by the
Friedmann--Lema\^itre--Robertson--Walker (FLRW) form, where $g_{00}(x)=1$,
and $g_{ij}(x)=-a(t)^{2}\delta_{ij}$ with $a(t)$ denoting the scale
factor of the three-dimensional space\footnote{As is customary in many inflationary scenarios including $\alpha$-attractors and the Starobinsky model, we adopt a fixed FLRW background throughout, ignoring back-reaction effects.}. The action, after dividing
by the spatial volume (while retaining the notation $S_{\phi}$),
becomes 
\begin{eqnarray}
S_{\phi}=\int dt\;a(t)^{3}\left\{ |\dot{\phi}(t)|^{2}-V(|\phi(t)|^{2})\right\} =\int dt\left\{ a(t)^{-3}|\Pi(t)|^{2}-a(t)^{3}V(|\phi(t)|^{2})\right\} ,
\end{eqnarray}
where the canonically conjugate momentum is defined as $\Pi^{\dagger}=a(t)^{3}\dot{\phi}$.

Suppose the scalar field undergoes a squeezing transformation described
by a conformal mapping,
\color{black}
\begin{equation}
\phi\to F_{n}(\phi;\theta),
\end{equation}

\begin{equation}
\dot{\phi}\to F'{}_{n}^{-1}(\phi;\theta)\dot{\phi},
\end{equation}
\color{black}
where $F(z)$ is a holomorphic function of a complex variable $z$.
This squeezing is a type of canonical transformation that preserves
the commutation (or Poisson bracket) relations: 
\begin{equation}
\{\phi'(t),\Pi'(t)^{\dagger}\}_{\text{PB}}=\{\phi(t),\Pi(t)^{\dagger}\}_{\text{PB}}=1,\;\{\phi'(t)^{\dagger},\Pi'(t)\}_{\text{PB}}=\{\phi(t)^{\dagger},\Pi(t)\}_{\text{PB}}=1.
\end{equation}

The canonical structure can be most easily understood by analogy with
quantum mechanics, where $\Pi^{\dagger}\propto\partial_{\phi}$ and
$\Pi\propto\partial_{\phi}^{\dagger}$. Classically, the transformation
can be generated by the function 
\begin{equation}
W(\phi,\Pi^{\dagger};\phi^{\dagger},\Pi')=F(\phi){\Pi'}^{\dagger}+F(\phi)^{\dagger}\Pi',
\end{equation}
which yields $\phi'=\frac{\partial W}{\partial(\Pi^{'\dagger})}$
and $\Pi^{\dagger}=\frac{\partial W}{\partial\phi}$.

Accordingly, we have
\begin{eqnarray}
\phi'=F(\phi),\ \Pi^{'\dagger}=F'(\phi)^{-1}\Pi^{\dagger}=F'(\phi)^{-1}a(t)^{3}\dot{\phi}.
\end{eqnarray}

The action after squeezing-induced effective transformation can thus be written
in terms of the original field $\phi$ as, 
\begin{eqnarray}
S'_{\phi}=\int dt\left\{ a(t)^{-3}|\Pi'(t)|^{2}-a(t)^{3}V(|\phi'(t)|^{2})\right\} =\int dt\;a(t)^{3}\left\{ \frac{|\dot{\phi}|^{2}}{|F'(\phi)|^{2}}-V(|F(\phi)|^{2})\right\} .\label{S'phi}
\end{eqnarray}

\color{black}
The above action should therefore be understood not as a symmetry-transformed copy of the original action, but as the effective action governing the classical configuration around the squeezed vacuum.
\color{black}

Next, we introduce a new scalar field $\chi(t)$, defined so that
the kinetic term becomes canonical. Specifically, we set 
\begin{eqnarray}
\dot{\chi}=\frac{\dot{\phi}}{F'(\phi)},\label{chi}
\end{eqnarray}
which determines $\chi$ as a function of $\phi$ via
\begin{eqnarray}
\chi(\phi)=\int_{0}^{\phi}\frac{d\phi}{F'(\phi)}.\label{chi(phi)}
\end{eqnarray}
We assume that $\phi$ can be inverted to give $\phi=\phi(\chi)$.
(We note that there is a phase ambiguity in the definition of $\chi$
(\ref{chi}), which may be fixed as required.)

The conformally transformed action can then be expressed in terms
of $\chi$ as 
\begin{eqnarray}
S'_{\chi}=S'_{\phi}=\int dt\;a(t)^{3}\left\{ |\dot{\chi}(t)|^{2}-V_{\chi}(\chi)\right\} ,
\end{eqnarray}
where the new potential, denoted as $V_{\chi}(\chi)$, is related
to the original potential by 
\begin{eqnarray}
V_{\chi}(\chi)=V(|F(\phi(\chi))|^{2}).
\end{eqnarray}

To construct a Virasoro-based inflation model, it is necessary to
explicitly specify the conformal transformation $F_{n}(\phi,\theta)$,
which corresponds to the $n$-th Virasoro squeezing with finite parameter
$\theta$. This transformation is defined by\footnote{The same analysis can be done for the momentum $\Pi$ \cite{Katagiri2019}.} 

\begin{eqnarray}
\phi'=F_{n}(\phi;\theta)\equiv\phi_{n}(\theta)=\hat{S}_{n}(\theta)\phi\hat{S}_{n}(\theta)^{-1}.
\end{eqnarray}

Differentiating both sides of this equation with respect to $\theta$
yields 
\begin{eqnarray}
\frac{d\phi_{n}(\theta)}{d\theta}=[\hat{L}_{n},\phi_{n}(\theta)]=\phi_{n}(\theta)^{n+1},\label{(21)}
\end{eqnarray}
where we have used $\hat{L}_{n}=\hat{S}_{n}(\theta)\hat{L}_{n}\hat{S}_{n}(\theta)^{-1}=\hat{L}_{n}(\theta)$.

Solving this differential equation, as introduced in Subsection 2.1,
we obtain:
\begin{eqnarray}
F_{n}(\phi;\theta)=\phi_{n}(\theta)=\frac{\phi}{\left(1-n\theta\phi^{n}\right)^{1/n}}.\label{def of Fn}
\end{eqnarray}

This allows us to construct a family of inflationary models parameterized
by the Virasoro mode $n$:
\begin{eqnarray}
S^{'(n)}=\int dt\;a(t)^{3}\left\{ |\dot{\chi}_{n}(t)|^{2}-V_{\chi}^{(n)}(\chi_{n})\right\} .
\end{eqnarray}

We focus on the case where the original potential takes the form $|\phi|^{m}$,
\begin{eqnarray}
V_{\chi}^{(m,n)}(\chi_{n})=\frac{\lambda_{m}}{m}|F_{n}(\phi(\chi_{n});\theta)|^{m},
\end{eqnarray}
where $\phi(\chi_{n})$ is the inverse function of $\chi_{n}(\phi)$,
defined as
\begin{eqnarray}
\chi_{n}(\phi)=\int_{0}^{\phi}d\phi\left(\frac{dF_{n}(\phi;\theta)}{d\phi}\right)^{-1}=\int_{0}^{\phi}d\phi\left(1-n\theta\phi^{n}\right)^{1+1/n}.\label{def of chi}
\end{eqnarray}

\color{black}
We should note here that a general Witt/Virasoro transformation is not restricted to a single mode. 
A general holomorphic vector field can be written in terms of an arbitrary function $\xi(\phi)$ as
\begin{equation}
\xi(\phi)\partial_\phi .
\end{equation}
Using the Witt/Virasoro generators, this vector field can be expanded as
\begin{equation}
\xi(\phi)\partial_\phi
=
\sum_n c_n \phi^{n+1}\partial_\phi
=
\sum_n c_n L_n .
\end{equation}
Therefore, a general squeezing transformation may be generated by a linear combination of several $L_n$ modes.

In the present paper, however, our purpose is to isolate how each elementary Virasoro mode affects the inflaton potential and the observables $(n_s,r)$. 
For this reason, we first restrict our analysis to a single mode $L_n$. 
This restriction does not exclude a general vector field, but should be regarded as a first step toward the more general case. 
It may also be interpreted as a single-mode approximation to a situation in which a particular nonlinear mode is dominant in the external driving process.

We also comment on the use of a monomial potential of degree $m$ before squeezing. 
A general potential can be expanded as a Taylor or Laurent series,
\begin{equation}
V(\phi)=\sum_m \frac{\lambda_m}{m} |\phi|^m .
\end{equation}
The potential $V\propto |\phi|^m$ considered in this paper should be understood as one basis term in such an expansion. 
Moreover, in a given field regime, the leading non-vanishing term or the dominant asymptotic term can often be approximated by a single power $m$. 
The analysis of general potentials involving several powers of $m$ will be left for future work.

Thus, the labels $n$ and $m$ are not arbitrary choices, but basis components appearing in the decomposition of a general transformation and a general potential, respectively. 
The analysis for each pair $(n,m)$ therefore provides a systematic first step toward understanding the general case.
\color{black}

\subsection{Asymptotic Behavior and Plateau Formation in $n$-th Virasoro Inflation}

In this subsection, we examine how $n$-th order Virasoro squeezing
leads to the formation of plateau structures in the inflaton potential,
depending on the strength of squeezing.

A key observation is that, as dictated by the maximum modulus theorem,
the potential after Virasoro squeezing achieves its maximal value
at the boundary of its domain. Specifically, the transformed potential
$V(|F_{n}(\text{\ensuremath{\phi}(\ensuremath{\chi}}_{n});\theta)|)$attains
its maximum in the asymptotic limit. We will demonstrate how, depending
on the value of the Virasoro index $n$, the inflaton potential can
indeed develop a plateau at infinity.

Depending on the value of $n$ , the resulting inflaton potentials
fall into different asymptotic classes. In particular, the cases $n>0$,
$n<-2$ and $n=0,-1,-2$ exhibit distinct structures.

In this subsection, we analyze the asymptotic behavior of the inflaton
potential $V(\chi_{n})$ induced by Virasoro transformations with
various indices $n$. The resulting potential may exhibit features
such as plateau structures or power-law growth, each having significant
implications for the inflationary dynamics. We classify the models
into physically meaningful categories and focus on the cases $n>0$,
which yield viable inflationary scenarios, and $n=-2$ , which permits
analytical treatment.

\subsubsection{Special Case: $n=-2$}

We now examine the special case of $n=-2$ which allows for an approximate
analytical representation of the transformation and the potential
at leading order.

For $n=-2$, the conformal transformation takes the form
\begin{eqnarray}
F_{-2}(\phi;\theta)=\pm\sqrt{\phi^{2}+2\theta}.
\end{eqnarray}
Accordingly, the canonical field $\chi_{-2}$ is given by 
\begin{eqnarray}
\pm\chi_{-2}=\int^{\phi}\frac{d\phi}{\phi}\sqrt{\phi^{2}+2\theta}=\sqrt{\phi^{2}+2\theta}-\sqrt{2\theta}\tanh^{-1}\sqrt{\frac{\phi^{2}+2\theta}{2\theta}}.\label{integral formula}
\end{eqnarray}
Note that the integration constant can shift $\chi_{-2}$ . Therefore,
without loss of generality, we omit the $\pm$signs in what follows
and treat $F_{-2}$ and $\chi_{-2}$ as real-valued functions. 

The above relation leads to

\begin{eqnarray}
\tanh\left(\frac{F_{-2}(\phi;\theta)}{\sqrt{2\theta}}-\frac{\chi_{-2}}{\sqrt{2\theta}}\right)=\frac{F_{-2}(\phi;\theta)}{\sqrt{2\theta}}.\label{pictorial figure}
\end{eqnarray}

In the case of $\theta<0$, replacing $[\tanh\to\tan]$ and setting
$\theta=-|\theta|$, we obtain 
\begin{eqnarray}
\tan\left(\frac{F_{-2}(\phi;-|\theta|)}{\sqrt{2|\theta|}}-\frac{\chi_{-2}}{\sqrt{2|\theta|}}\right)=\frac{F_{-2}(\phi;-|\theta|)}{\sqrt{2|\theta|}}.
\end{eqnarray}

Coming back to the case of $\theta>0$, we find the following approximate
behavior: 
\begin{eqnarray}
F_{-2}(\phi;\theta)\sim\begin{cases}
~-\sqrt{2\theta},~\text{for}~\chi_{-2}\gg\sqrt{2\theta},\\
~~~~~0,~~~~~\text{for}~{\color{black}{\color{black}|\chi_{-2}|}\mathrel{\color{black}\ll}{\color{black}\sqrt{2\theta}}\mathpunct{\color{black},}}
\end{cases}
\end{eqnarray}
we find the asymptotic behavior 
\begin{eqnarray}
 &  & V_{\chi}^{(m,-2)}(\chi_{-2})=\frac{\lambda_{m}}{m}\left|F_{-2}(\chi_{-2})\right|^{m}\\
 &  & \sim\frac{\lambda_{m}}{m}\begin{cases}
~|2\theta|^{m/2}~\text{(giving plateau)}~~~~\text{\ensuremath{\mathrm{for}}}~~|\chi_{-2}|\gg|\sqrt{2\theta}|,\\
~0,\ \ \ \mathrm{for}~~|\chi_{-2}|\ll|\sqrt{2\theta}|.
\end{cases}
\end{eqnarray}

In contrast, for $\theta<0$, the behavior becomes
\begin{eqnarray}
F_{-2}(\phi;\theta)\sim\begin{cases}
~~~\chi_{-2},~~~\;\text{for}~\chi_{-2}\gg\sqrt{2|\theta|},\\
~{\color{black}\mathinner{\color{black}\frac{\pi}{2}}{\color{black}\sqrt{2|\theta|}}\mathpunct{\color{black},}{\color{black}~\text{for}~|\chi_{-2}|\ll\sqrt{2|\theta|}.}}
\end{cases}
\end{eqnarray}
The corresponding inflaton potential for $\theta<0$ is asymptotically

\begin{equation}
V_{\chi}^{(m,-2)}(\chi_{-2})(\theta<0)=\frac{\lambda_{m}}{m}\left|\chi_{2}\right|^{m}\ \text{for}\ |\chi_{-2}|\gg|\sqrt{2\theta}|,
\end{equation}
which resembles the original unsqueezed potential and is therefore
not of particular interest.

It is characteristic and interesting that for $\theta>0$, the potential
exhibits a plateau in the limit $|\chi_{-2}|\gg|\sqrt{2\theta}|$,
making it a viable inflationary scenario.

The analytical properties of the potential in this case allow for
a detailed discussion, which will be presented in Section 3.

\subsubsection{Model with $n>0$}

From the definition of the inflaton field $\chi_{n}$ in (\ref{def of chi}),
\begin{eqnarray}
\chi_{n}(\phi)=\int_{0}^{\phi}d\phi\left(1-n\theta\phi^{n}\right)^{1+1/n},
\end{eqnarray}
we observe that for $n>0$, the integral is dominated by large value
of $|\phi|$, leading to the approximation 
\begin{eqnarray}
\chi_{n}\approx\frac{(-n\theta)^{1+1/n}}{n+2}\phi^{n+2}~\text{for}~~n>0.
\end{eqnarray}

Furthermore, according to the maximum modulus theorem in complex analysis,
it is naturally understood that the modulus of the potential gives
a plateau (maximum) at $|\phi|\to\infty$, 
\begin{eqnarray}
|V_{m}(|\phi|^{2})|=\frac{\lambda_{m}}{m}\left|\frac{\phi}{(1-n\theta\phi^{n})^{1/n}}\right|^{m}\to\frac{\lambda_{m}}{m}|n\theta|^{-1/n},~\text{(plateau)}~\text{for}~|\phi|\to\infty.
\end{eqnarray}

Since $|\phi|\to\infty$ corresponds to $|\chi_{n}|\to\infty$ for
$n>0$, and $\chi_{n}$ and $\phi$ are related in power-law fashion,
the new potential $V_{\chi}^{(m,n)}(\chi_{n})$ also develops a plateau
at $|\chi_{n}|=\infty$, from which the inflaton field $\chi_{n}$
rolls down slowly. This naturally leads to viable slow-roll inflation
scenarios.

More explicitly, the leading-order relation between $\phi$ and $\chi_{n}$
is given by
\begin{eqnarray}
\phi\approx A_{n}\chi_{n}^{\frac{1}{n+2}},
\end{eqnarray}
where we have introduced 
\begin{eqnarray}
A_{n}\equiv\left(\frac{n+2}{(-n\theta)^{1+1/n}}\right)^{\frac{1}{n+2}}.\label{An}
\end{eqnarray}

Therefore, the inflaton potential for $n>0$, can be approximated
by, 
\begin{eqnarray}
V_{\chi}^{(m,n)}(\chi_{n})\approx\frac{\lambda_{m}}{m}\left|\frac{A_{n}\chi_{n}^{\frac{1}{n+2}}}{\left(1-n\theta(A_{n})^{n}\chi_{n}^{\frac{n}{n+2}}\right)^{1/n}}\right|^{m}.\label{n>0}
\end{eqnarray}

Interestingly, the inflaton potential $V_{\chi}^{(m,n)}(\chi_{n})$
for $n<-2$ takes the same formal expression as that for $n>0$; the
only difference is $n$ is negative, i.e., $n=-|n|$. However, the
asymptotic behavior differs, as the plateau emerges in the opposite
limit $\phi\to0$ rather than $\phi\to\infty$.

On the other hand, $n=0$ and $n=-1$ correspond to trivial transformations
of the potential.
\color{black}
These correspond to special cases of the squeezing transformation that act trivially on the potential: $F_0(\phi)$ corresponds to a simple rescaling, $F_{0}(\phi)=e^{\theta}\phi$, and $F_{-1}(\phi)=\phi+\theta$ corresponds to a constant shift. In this sense, the $n=-1$ case may serve as a simple diagnostic of shift symmetry: if the original potential satisfies $V(\phi+\theta)=V(\phi)$, it remains invariant under this transformation. Since such cases do not generate a new plateau structure in the class of models considered here, we do not pursue them further.
\color{black}

The numerical and approximate analytical results for the case $n>0$
will be discussed in detail in Section 4.

\subsection{Analytical Origin of the Plateau via the Maximum Modulus Theorem}

In this section, we provide a complex analysis perspective to explain
why $n$-th order Virasoro squeezing generates plateau-like potentials,
as discussed in the previous section.

The mathematical underpinning of this phenomenon is the classical
maximum modulus theorem in complex analysis\cite{Ahlfors1979}. We
summarize the theorem below.
\begin{prop*}
Maximum Modulus Theorem

Let $f(z)$ be a holomorphic (analytic) and non-constant function
inside and on a closed curve $C$. Then, the modulus $|f(z)|$ does
not attain its maximum value anywhere in the interior of $C$.
\end{prop*}
This result can be proven using Cauchy's integral theorem and the
triangle inequality. From this theorem, the following corollary immediately
follows.
\begin{cor*}
The maximum of $|f(z)|$ is always attained on the boundary $C$.
That is, by the maximum modulus theorem, $|f(z)|$ cannot have a maximum
in the interior of $C$. However, by the Weierstrass theorem, any
continuous function must attain its maximum somewhere, and therefore
the maximum must occur on the boundary $C$.
\end{cor*}
Therefore, the maximum value of a holomorphic function always appears
on the boundary of its domain. When the potential is expressed as
a transformation of a holomorphic function, its asymptotic behavior---i.e.,
its value at the boundary---becomes dominant. This underlies the
emergence of plateau-like structures in the transformed inflaton potential.

Furthermore, by the Cauchy integral formula (Goursat's theorem), the
first derivative of $f(z)$ at $z=a$ can be written as
\begin{equation}
f^{1}(a)=\frac{1}{2\pi i}\oint_{C}\frac{f(z)}{(z-a)^{2}}dz.
\end{equation}
Letting $z-a=re^{i\theta}$, we obtain the inequality

\begin{equation}
|f^{1}(a)|\leq\frac{1}{2\pi r}\int_{0}^{2\pi}\left|f(a+re^{i\theta})\right|d\theta.
\end{equation}
If we assume the boundary is at infinity and that the boundary value
of $f(z)$ satisfies $\lim_{|z|\to\infty}f(z)=f_{\infty}$(a finite
value), then
\begin{equation}
\lim_{|z|\to\infty}|f^{1}(z)|\leq\frac{1}{r}|f_{\infty}|.
\end{equation}
Thus, if $f_{\infty}=0$, the derivative vanishes asymptotically and
a plateau is formed.

\subsubsection{Plateau Formation via Virasoro Squeezing}

The Virasoro squeezing transformation is given by
\begin{equation}
F_{n}(\phi;\theta)=\frac{\phi}{(1-n\theta\phi^{n})^{1/n}}.
\end{equation}
In the limit where $\phi$ becomes large, the transformation asymptotically
approaches
\begin{equation}
F_{n}(\phi;\theta)\approx\frac{1}{(-)^{1/n}n^{1/n}\theta^{1/n}}.
\end{equation}
The potential after transformation takes the form
\begin{equation}
V_{m}(|F_{n}(\phi;\theta)|)=\frac{\lambda}{m}\left(\frac{\phi}{(1-n\theta\phi^{n})^{1/n}}\right)^{m}.
\end{equation}

\color{black}
For $n>0$, the plateau behavior is obtained in the asymptotic regime when the branch and sign of $\theta$ are chosen so that the transformation remains real and regular along the inflationary trajectory. In the numerical analysis below, this corresponds to the regular branch with $\theta<0$.
\color{black}

\subsubsection*{Analysis in the $\chi$ Representation }

In the $\chi$ representation, the relationship between $\phi$ and
$\chi_{n}$ can be approximated as

\begin{equation}
\phi(\chi_{n})\approx A_{n}\chi_{n}^{1/(n+2)}
\end{equation}
where $A_{n}$ is a constant determined by the transformation.

For $n>-2$, both $\chi$ and $\phi$ become large in the asymptotic
regime, and as a result, $F_{n}(\phi,\theta)$ forms a plateau at
large field values. Conversely, for $n<-2$, the limit $\chi_{n}\to0$
corresponds to $\phi\to\infty$, so $F_{n}(\phi,\theta)$ develops
a plateau in the vicinity of $\chi_{n}\to0$. 

Thus, the asymptotic behavior of the potential in our model can be
understood as a generic feature of holomorphic transformations, as
established by complex analysis.

\section{The $n=-2$ Virasoro Squeezed model}

In this section, we analyze the case of $n=-2$ which admits an approximate
analytical treatment at leading order. We show that, to leading order,
the resulting potential acquires a hyperbolic tangent ($\tanh$) form,
which reproduces the well-known Starobinsky potential---or, equivalently,
the $t$-model potential in the $\alpha$-attractor class.

Crucially, these results are obtained by applying $n=-2$ Virasoro
squeezing to a potential of the form $V=\frac{\lambda}{m}|\phi|^{m}$,
without introducing any special coupling to gravity. We also demonstrate
that, at higher orders of approximation, the $n=-2$ model exhibits
deviations from the $t$-model potential.

Near the plateau, in the limit $|\chi_{-2}|\approx\infty$, the relation
between $F_{-2}$ and $\chi_{-2}$ can be approximated. From the defining
relation (see (\ref{pictorial figure})), we use the following approximation

\begin{equation}
F_{-2}(\phi;\theta)=\sqrt{2\theta}\tanh\left(\frac{F_{-2}(\phi;\theta)}{\sqrt{2\theta}}-\frac{\chi_{-2}}{\sqrt{2\theta}}\right)\approx\varepsilon\sqrt{2\theta}\tanh\left(-\frac{1}{\sqrt{2\theta}}|\chi_{-2}|\right),\ \varepsilon=\chi_{2}/|\chi_{-2}|,\label{eq:Fnminu2}
\end{equation}
where we have set $\varepsilon=1$ without loss of generality.

This leads to
\begin{equation}
V_{\chi,0}^{(-2)}=V_{\chi}^{(-2)}(\sqrt{2\theta}\tanh\left(-\frac{1}{\sqrt{2\theta}}|\chi_{-2}|\right)\approx V_{\chi}^{(-2)}\left(-\sqrt{2\theta}\left(1-2e^{-\frac{2|\chi_{-2}|}{\sqrt{2\theta}}}\right)\right).
\end{equation}
It should be emphasized that the original potential $V_{\chi}^{(-2)}(\phi)$
before squeezing can be arbitrary; thus, the Starobinsky-type and
$\alpha$-attractor-type $t$-models emerge as lowest-order approximations
within this framework.
\color{black}
However, as will be discussed below, this is not an exact t-model under higher-order corrections. Thus, even in the \(n=-2\) case, the model is not exactly identical to the conventional \(t\)-model beyond the leading-order approximation.
\color{black}

Incorporating the next-order correction, we set
\begin{equation}
F_{-2}=-\sqrt{2\theta}\tanh(\frac{\chi_{-2}}{\sqrt{2\theta}})+\Delta F_{-2}.
\end{equation}
Substituting this expression and expanding to leading order, we obtain
\begin{equation}
-\sqrt{2\theta}\tanh(\frac{\chi_{-2}}{\sqrt{2\theta}})+\Delta F_{-2}\approx-\sqrt{2\theta}\tanh(\frac{\chi_{-2}}{\sqrt{2\theta}})-\left(1-\tanh^{2}\frac{\chi_{-2}}{\sqrt{2\theta}}\right)\left(1+\frac{1}{\sqrt{2\theta}}\Delta F_{-2}\right).
\end{equation}
Solving for $\Delta F_{-2}$, we find
\begin{equation}
\Delta F_{-2}=-\frac{2\sqrt{2\theta}}{3+\cosh[2\frac{\chi_{-2}}{\sqrt{2\theta}}]}.
\end{equation}
Thus, the corrected expression for $F_{-2}$ is
\begin{equation}
F_{-2}\approx-\sqrt{2\theta}\tanh(\frac{\chi_{-2}}{\sqrt{2\theta}})-\frac{2\sqrt{2\theta}}{3+\cosh[2\frac{\chi_{-2}}{\sqrt{2\theta}}]}.
\end{equation}
Accordingly, the inflaton potential becomes
\begin{equation}
V_{\chi}=V_{\chi}^{(-2)}\left(\sqrt{2\theta}\tanh\left(-\frac{1}{\sqrt{2\theta}}|\chi_{-2}|\right)-\frac{2\sqrt{2\theta}}{3+\cosh[2\frac{\chi_{-2}}{\sqrt{2\theta}}]}\right).
\end{equation}
This shows that the $n=-2$ model introduces corrections to the $\alpha$-attractor
$t$-model potential at higher orders of approximation.

For a potential of the form $V^{(m)}=\frac{\lambda_{m}}{m}\phi^{m}$,
the inflaton potential is approximately given by

\begin{eqnarray}
V_{\chi}^{(m,-2)}=\frac{\lambda_{m}}{m}\varepsilon^{m}\left(\sqrt{2\theta}\right)^{m}\left(\tanh\left(-\frac{1}{\sqrt{2\theta}}|\chi_{-2}|\right)+O(\Delta F_{-2})\right)^{m}{\color{black}{\color{black}\label{potentialforn=-2}.}}
\end{eqnarray}
The slow-roll parameters \(\epsilon_V\) and \(\eta_V\) are computed numerically, and the observables \(n_s\) and \(r\) are then obtained from them.\footnote{\color{black}The quantities \(n_s\) and \(r\) are not slow-roll parameters themselves, but observable quantities derived from the slow-roll parameters \(\epsilon_V\) and \(\eta_V\). We use them as standard indicators of scalar and tensor primordial perturbations.\color{black}} For example, setting
$m=2$, $\theta=1$, and $N_{e}=60$, we obtain $n_{s}=0.966837$
and $r=0.00109071$, which are within the Planck observational constraints\cite{PLACNK2018, Ade2021}.

\begin{figure}
\centering
\includegraphics[width=1\textwidth,totalheight=1\textheight,keepaspectratio]{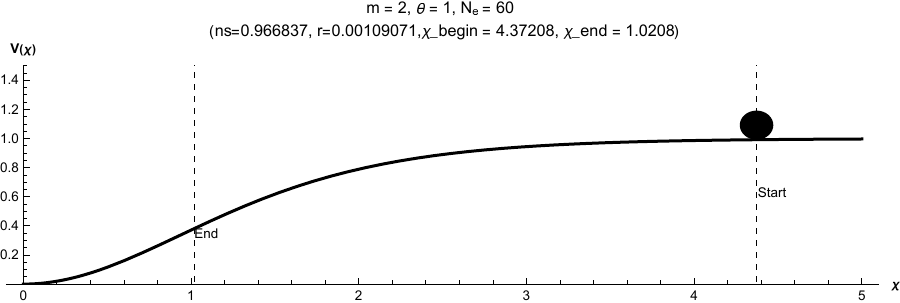}

\caption{Potential $V(\chi)$ for $n=-2,m=2,\theta=1$. The horizontal axis
represents the inflaton field $\chi$, and the vertical axis shows
the potential $V(\chi)$. The vertical dashed lines indicate the beginning
and end points of the slow-roll inflation phase.\label{fig:Potentialn-2}}
\end{figure}

Figure \ref{fig:Potentialn-2} shows the potential for $n=-2,\ m=2,$
and $\theta=1$, along with the starting and ending points of the
slow-roll phase and the evolution of the inflaton field.

\begin{figure}
\centering
\includegraphics[width=0.5\textwidth,totalheight=0.3\textheight]{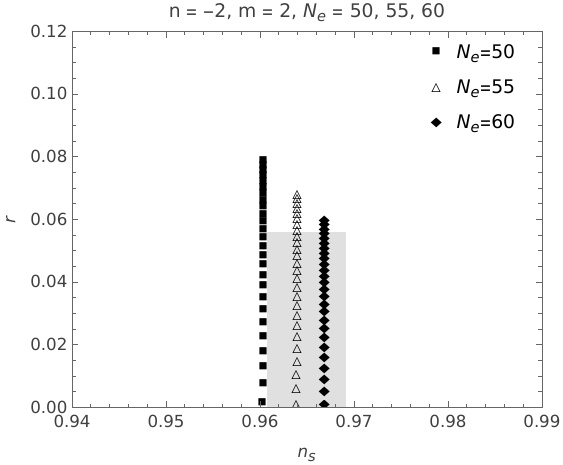}

\caption{$(n_{s},r)$ parametric plot with $\theta=0\sim100$, $n=-2,m=2,N_{e}=50,55,60$.
The gray shaded region indicates the range allowed by current observational
constraints.} \label{fig:minus2parametric-prot-efoldings}
\end{figure}

\begin{figure}
\includegraphics[width=0.5\textwidth,totalheight=0.3\textheight]{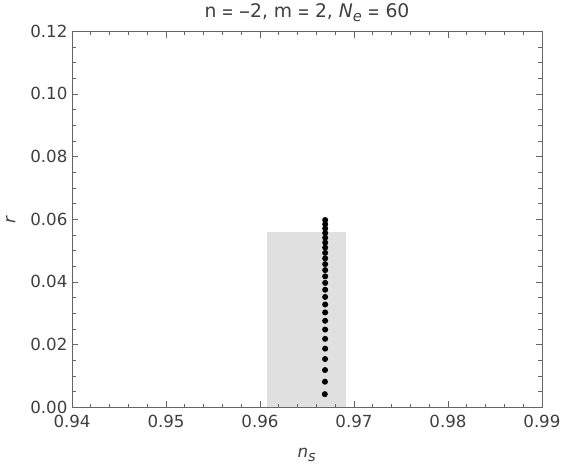}\includegraphics[width=0.5\textwidth,totalheight=0.3\textheight]{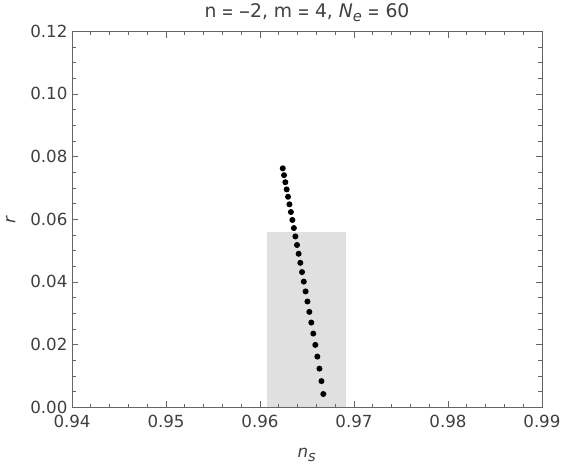}\\
\includegraphics[width=0.5\textwidth,totalheight=0.3\textheight]{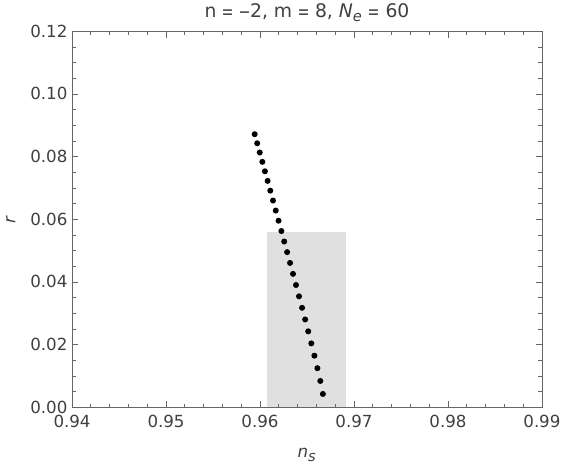}

\caption{$(n_{s},r)$ parametric plot with $\theta=0\sim100$, $n=-2,m=\text{2,\ 4,\ 8},N_{e}=60$.
The gray shaded region indicates the range allowed by current observational
constraints.} \label{fig:minus2parametric-prot-variant-m}
\end{figure}

Figure \ref{fig:minus2parametric-prot-efoldings} illustrates the
behavior of $(n_{s},r)$ in the parameter space $\theta=0\sim100$
for $N_{e}=50,55,60$. The gray shaded region indicates the range
allowed by current observational constraints. These results demonstrate
that the full range of observed values can be covered by adjusting
the e-folding number. Figure \ref{fig:minus2parametric-prot-variant-m}
shows the behavior of $(n_{s},r)$ for varying $m=2,4,8$, again as
a function of $\theta$ in the range $0\sim100$. As $m$ increases,
the vertical trajectories tilt diagonally, while still remaining consistent
with observational bounds.

While the leading-order behavior of the potential can be captured
analytically, a precise evaluation of the observables $n_{s}$
and $r$ requires numerical computation. Therefore, numerical analysis
is essential for fully assessing the model's viability.

These results confirm that, in the $n=-2$ Virasoro squeezed model,
the inflaton potential reproduces the essential features of the Starobinsky
and $\alpha$-attractor $t$-model potentials at leading order, while
allowing for characteristic corrections at higher orders.
\color{black}
For other Virasoro modes n, the transformed potentials and the resulting observables $(n_s,r)$ differ from the leading-order $n=-2$ attractor limit.
\color{black}
 In the next section, we generalize our analysis to other values of $n$ and examine
the resulting phenomenology through numerical calculations.

\section{Numerical Results and Discussion}

Although the case $n=-2$ admits an analytic expression for the conformal
transformation, the evaluation of the observables $n_{s}$
and $r$ relies on the approximate expression (\ref{eq:Fnminu2})
and its numerical integration. This semi-analytic treatment was already
presented in Section 3.

We now turn to other values of $n$, for which no closed-form expression
is available for the transformation $F_{n}(\phi;\theta)$, and a fully
numerical approach is required.

As discussed in Section 3, the inflationary potential generated by
Virasoro squeezing satisfies the slow-roll conditions. To obtain precise
predictions for the observables $n_{s}$ and $r$, we now adopt a
numerical approach. We describe the computational setup and algorithm,
and compare the resulting predictions with observational constraints.
For convenience, we define the transformed field variable $y\equiv F_{n}(\phi)$,
which will be used in the following analysis.

Unlike the approximate analysis in Section 3, we here rely entirely
on numerical methods. Instead of using the standard e-folding expression
involving $V/V'$, we determine the endpoint $y_{e}$ by solving $\epsilon_{V}(y_{e})=1$,
and then obtain the initial value $y_{b}$ by integrating the e-folding
number numerically. The slow-roll parameters $\epsilon_{V}$ and $\eta_{V}$
at $y_{b}$ are then used to compute $n_{s}$ and $r$\footnote{Definitions of the slow-roll parameters $\epsilon_{V}$ and $\eta_{V}$
are provided in Appendix A.}. 

This approach enables us to systematically treat the different branches
of the Virasoro transformation, including nontrivial root structures,
and to explore regions of parameter space that are inaccessible to
analytical approximations.

\subsection{Numerical Setup}

The technical details of our numerical procedure are given in Appendix
A. Here, we briefly outline the main computational setup, with particular
emphasis on the treatment of the branch structures that arise from
the Virasoro-generated conformal transformations.

The transformation $\phi\to F_{n}(\phi;\theta)$ involves fractional
powers (i.e., $n$-th roots), and the evaluation of these roots depends
on the proper choice of complex branch\footnote{Strictly speaking, the conformally transformed field $F_{n}(\phi)$
and its canonically normalized counterpart $\chi_{n}$ are multivalued
complex functions. A fully consistent treatment would require formulating
slow-roll inflation as a multi-component field theory, where the inflaton
trajectory follows a curved path in field space. This introduces significant
complications in analyzing the dynamics and evaluating the potential.
In this work, we simplify the analysis by restricting to real-valued
field configurations and explicitly accounting for branch structures
arising from root ambiguities. A more complete treatment that incorporates
complex inflaton trajectories\cite{Wands2002} is left for future
investigation.}. In our analysis, we define and classify three representative branches
according to the sign of the Virasoro index $n$ and the manner in
which the root is taken:
\begin{itemize}
\item {[}Branch1+{]} (for $n>0)$ 
\end{itemize}
This branch corresponds to choosing a non-principal branch of the
$n$-th root for the transformation variable. It was introduced to
investigate the behavior near singular limits, such as $\theta\to0$.
\begin{itemize}
\item {[}Branch2+{]}(for $n>0)$ 
\end{itemize}
In this standard branch, the principal $n$-th root is adopted. This
leads to well-behaved potentials and is considered the most physically
natural choice for $n>0$.
\begin{itemize}
\item {[}Branch3-{]}(for $n<0)$ 
\end{itemize}
In this case, the transformation involves an $|n|$-th root rather
than an $n$-th root. With the variable change $t=\frac{1}{|n\theta|}\phi^{|n|}$,
the transformation is given by $y=F(\phi)=(t+\mathrm{sign}(\theta))^{1/|n|}$
.

Among these, Branch 2+ and, in some cases, Branch 1+ yield potentials
that exhibit plateau-like behavior after transformation. As a result,
these branches naturally satisfy the slow-roll conditions and provide
viable predictions for the spectral index $n_{s}$ and the tensor-to-scalar
ratio $r$ within certain regions of parameter space. Also, Branch
3\textminus{} does yield valid parameter sets that are consistent
with observational bounds within the scanned range for $n=-2$ which
is consistent with Analytical case.

Therefore, in the following subsections, we focus primarily on Branch
2+ and Branch 1+ and Branch 3- in our numerical analysis.

\subsection{Numerical Results}

We performed a parameter scan for each branch, exploring $n=+1\sim+6$
(and $n=-1\sim-6$ for Branch 3), $m=-16\sim+16$ and $\theta\in[-10^{7},10^{-2}]\cup[10^{-2},10^{7}]$,
in order to identify combinations of $(n,m)$ that yield predictions
consistent with observational bounds. \color{black}Here, we used a value of $60$ for $N_e$\color{black}.
Representative results for each
branch are summarized in Table \ref{tab:Branch-1} -  \ref{tab:Branch-3},
respectively. For each branch, we list the Virasoro index $n$, the
potential power $m$, the minimum and maximum values of $\theta(\theta_{\mathrm{min}},\theta_{\mathrm{max}})$
for which the predictions agree with observational constraints, as
well as the resulting spectral index $n_{s}$ and tensor-to-scalar
ratio $r$, together with their corresponding uncertainties $(\text{\ensuremath{\sigma}}_{n_{s}},\text{\ensuremath{\sigma}}_{r}$).
For each pair $(n,m)$, the standard deviations $\sigma_{n_{s}}$\LyXZeroWidthSpace{}
and $\sigma_{r}$\LyXZeroWidthSpace{} are computed from all solutions
obtained by scanning over a range of values of the squeezing parameter
$\theta$, including cases where multiple valid solutions exist for
the same $\theta$. These values quantify the predictive variation
of the model under parameter scanning and should not be interpreted
as observational uncertainties.

\color{black}
The procedure used to construct this set of solutions is as follows. 
For fixed $(n,m)$ and for each branch, we scan the squeezing parameter $\theta$ over a prescribed range. 
For each value of $\theta$, we determine the end point of inflation by solving $\epsilon_V=1$, and then determine the field value at horizon exit from the e-folding condition with $N_e=60$. 
The slow-roll parameters $\epsilon_V$ and $\eta_V$ are evaluated at this point, and the corresponding observables $(n_s,r)$ are computed. 
We retain the solutions that are real, regular on the chosen branch, and consistent with the observational bounds. 
When more than one valid branch or root gives an allowed solution for the same value of $\theta$, all such solutions are included as separate entries in the statistical sample. 
The standard deviations $\sigma_{n_s}$ and $\sigma_r$ are then computed from this set of accepted solutions. 
Further details of this procedure are given in Appendix~A.3.
\color{black}

\begin{table}
\begin{tabular}{|c|c|c|c|c|c|c|c|}
\hline 
$n$ & $m$ & $\theta_{\mathrm{min}}$ & $\theta_{\mathrm{max}}$ & $n_{s}$ & $\sigma_{n_{s}}$ & $r$ & $\sigma_{r}$\tabularnewline
\hline 
\hline 
4 & 0.5 & $-1.17\times 10^{-4}$ & $1.17\times 10^{-4}$ & 0.96885 & $1.3\times 10^{-4}$ & 0.01248 & $1.9\times 10^{-3}$\tabularnewline
\hline 
4 & 1 & $-5.46\times 10^{-5}$ & $5.46\times 10^{-5}$ & 0.96771 & $6.2\times 10^{-4}$ & 0.02818 & $8.7\times 10^{-3}$\tabularnewline
\hline 
4 & 2 & $-2.73\times 10^{-5}$ & $2.73\times 10^{-5}$ & 0.96635 & $1.6\times 10^{-3}$ & 0.03658 & $8.9\times 10^{-3}$\tabularnewline
\hline 
4 & 3 & $-1.93\times 10^{-5}$ & $1.93\times 10^{-5}$ & 0.96651 & $1.5\times 10^{-3}$ & 0.04143 & $7.3\times 10^{-3}$\tabularnewline
\hline 
4 & 4 & $-1.68\times 10^{-5}$ & $1.68\times 10^{-5}$ & 0.96718 & $1.3\times 10^{-3}$ & 0.04368 & $6.1\times 10^{-3}$\tabularnewline
\hline 
4 & 5 & $-1.46\times 10^{-5}$ & $1.46\times 10^{-5}$ & 0.96740 & $1.1\times 10^{-3}$ & 0.04724 & $5.1\times 10^{-3}$\tabularnewline
\hline 
4 & 6 & $-1.37\times 10^{-5}$ & $1.37\times 10^{-5}$ & 0.96788 & $7.5\times 10^{-4}$ & 0.04893 & $3.6\times 10^{-3}$\tabularnewline
\hline 
4 & 7 & $-1.37\times 10^{-5}$ & $1.37\times 10^{-5}$ & 0.96813 & $5.7\times 10^{-4}$ & 0.05109 & $2.8\times 10^{-3}$\tabularnewline
\hline 
4 & 8 & $-1.37\times 10^{-5}$ & $1.37\times 10^{-5}$ & 0.96846 & $3.9\times 10^{-4}$ & 0.05235 & $2.0\times 10^{-3}$\tabularnewline
\hline 
4 & 9 & $-1.37\times 10^{-5}$ & $1.37\times 10^{-5}$ & 0.96868 & $2.5\times 10^{-4}$ & 0.05375 & $1.3\times 10^{-3}$\tabularnewline
\hline 
4 & 10 & $-1.46\times 10^{-5}$ & $1.46\times 10^{-5}$ & 0.96891 & $1.5\times 10^{-4}$ & 0.05470 & $7.9\times 10^{-4}$\tabularnewline
\hline 
\end{tabular}

\caption{Representative parameter sets for Branch 1. Listed are the Virasoro
index $n$, the potential power $m$, the allowed range of the squeezing
parameter $\theta$, the predicted values of the spectral index $n_{s}$
and tensor-to-scalar ratio $r$, and their respective standard deviations.
The standard deviations ($\sigma_{n_{s}},\sigma_{r}$) are computed
over all valid values of $\theta$ that satisfy the observational
constraints for each $(n,m)$ pair.}
\label{tab:Branch-1}
\end{table}

\begin{table}
\begin{tabular}{|c|c|c|c|c|c|c|c|}
\hline 
$n$ & $m$ & $\theta_{\mathrm{min}}$ & $\theta_{\mathrm{max}}$ & $n_{s}$ & $\sigma_{n_{s}}$ & $r$ & $\sigma_{r}$\tabularnewline
\hline 
\hline 
3 & 2 & $-1.65\times 10^{-4}$ & $-6.72\times 10^{-5}$ & 0.96807 & $6.5\times 10^{-4}$ & 0.04588 & $5.3\times 10^{-3}$\tabularnewline
\hline 
3 & 3 & $-1.90\times 10^{-4}$ & $-1.02\times 10^{-4}$ & 0.96816 & $6.0\times 10^{-4}$ & 0.04880 & $3.6\times 10^{-3}$\tabularnewline
\hline 
3 & 4 & $-2.03\times 10^{-4}$ & $-1.44\times 10^{-4}$ & 0.96850 & $4.1\times 10^{-4}$ & 0.05120 & $2.3\times 10^{-3}$\tabularnewline
\hline 
3 & 5 & $-2.18\times 10^{-4}$ & $-1.90\times 10^{-4}$ & 0.96872 & $2.0\times 10^{-4}$ & 0.05386 & $1.1\times 10^{-3}$\tabularnewline
\hline 
4 & 0.5 & $-1.17\times 10^{-4}$ & $-3.36\times 10^{-5}$ & 0.96885 & $1.3\times 10^{-4}$ & 0.01248 & $1.9\times 10^{-3}$\tabularnewline
\hline 
4 & 1 & $-5.46\times 10^{-5}$ & $-2.78\times 10^{-6}$ & 0.96771 & $6.3\times 10^{-4}$ & 0.02818 & $8.8\times 10^{-3}$\tabularnewline
\hline 
4 & 2 & $-2.73\times 10^{-5}$ & $-2.98\times 10^{-6}$ & 0.96635 & $1.6\times 10^{-3}$ & 0.03658 & $8.9\times 10^{-3}$\tabularnewline
\hline 
4 & 3 & $-1.93\times 10^{-5}$ & $-3.66\times 10^{-6}$ & 0.96651 & $1.5\times 10^{-3}$ & 0.04143 & $7.3\times 10^{-3}$\tabularnewline
\hline 
4 & 4 & $-1.68\times 10^{-5}$ & $-4.51\times 10^{-6}$ & 0.96718 & $1.3\times 10^{-3}$ & 0.04368 & $6.2\times 10^{-3}$\tabularnewline
\hline 
4 & 5 & $-1.46\times 10^{-5}$ & $-5.18\times 10^{-6}$ & 0.96740 & $1.1\times 10^{-3}$ & 0.04724 & $5.1\times 10^{-3}$\tabularnewline
\hline 
4 & 6 & $-1.37\times 10^{-5}$ & $-6.38\times 10^{-6}$ & 0.96788 & $7.5\times 10^{-4}$ & 0.04893 & $3.6\times 10^{-3}$\tabularnewline
\hline 
4 & 7 & $-1.37\times 10^{-5}$ & $-7.32\times 10^{-6}$ & 0.96813 & $5.8\times 10^{-4}$ & 0.05109 & $2.8\times 10^{-3}$\tabularnewline
\hline 
4 & 8 & $-1.37\times 10^{-5}$ & $-9.01\times 10^{-6}$ & 0.96846 & $3.9\times 10^{-4}$ & 0.05235 & $2.0\times 10^{-3}$\tabularnewline
\hline 
4 & 9 & $-1.37\times 10^{-5}$ & $-1.04\times 10^{-5}$ & 0.96868 & $2.5\times 10^{-4}$ & 0.05375 & $1.3\times 10^{-3}$\tabularnewline
\hline 
4 & 10 & $-1.46\times 10^{-5}$ & $-1.27\times 10^{-5}$ & 0.96891 & $1.5\times 10^{-4}$ & 0.05470 & $8.3\times 10^{-4}$\tabularnewline
\hline 
5 & 0.5 & $-6.72\times 10^{-5}$ & $-1.30\times 10^{-6}$ & 0.96675 & $1.2\times 10^{-3}$ & 0.01251 & $5.3\times 10^{-3}$\tabularnewline
\hline 
5 & 1 & $-1.57\times 10^{-5}$ & $-1.07\times 10^{-7}$ & 0.96603 & $1.8\times 10^{-3}$ & 0.02598 & $1.4\times 10^{-2}$\tabularnewline
\hline 
5 & 2 & $-4.21\times 10^{-6}$ & $-1.86\times 10^{-7}$ & 0.96479 & $2.5\times 10^{-3}$ & 0.02995 & $9.8\times 10^{-3}$\tabularnewline
\hline 
5 & 3 & $-2.11\times 10^{-6}$ & $-1.62\times 10^{-7}$ & 0.96543 & $2.5\times 10^{-3}$ & 0.03414 & $9.3\times 10^{-3}$\tabularnewline
\hline 
5 & 4 & $-1.49\times 10^{-6}$ & $-1.52\times 10^{-7}$ & 0.96582 & $2.2\times 10^{-3}$ & 0.03826 & $8.2\times 10^{-3}$\tabularnewline
\hline 
5 & 5 & $-1.05\times 10^{-6}$ & $-1.62\times 10^{-7}$ & 0.96613 & $1.8\times 10^{-3}$ & 0.04189 & $6.9\times 10^{-3}$\tabularnewline
\hline 
5 & 6 & $-9.17\times 10^{-7}$ & $-1.74\times 10^{-7}$ & 0.96667 & $1.6\times 10^{-3}$ & 0.04396 & $6.3\times 10^{-3}$\tabularnewline
\hline 
5 & 7 & $-7.45\times 10^{-7}$ & $-1.86\times 10^{-7}$ & 0.96699 & $1.3\times 10^{-3}$ & 0.04631 & $5.4\times 10^{-3}$\tabularnewline
\hline 
5 & 8 & $-6.95\times 10^{-7}$ & $-2.14\times 10^{-7}$ & 0.96766 & $9.5\times 10^{-4}$ & 0.04666 & $4.1\times 10^{-3}$\tabularnewline
\hline 
5 & 9 & $-6.49\times 10^{-7}$ & $-2.30\times 10^{-7}$ & 0.96774 & $8.1\times 10^{-4}$ & 0.04916 & $3.6\times 10^{-3}$\tabularnewline
\hline 
5 & 10 & $-6.05\times 10^{-7}$ & $-2.64\times 10^{-7}$ & 0.96796 & $6.9\times 10^{-4}$ & 0.05071 & $3.2\times 10^{-3}$\tabularnewline
\hline 
5 & 11 & $-6.05\times 10^{-7}$ & $-3.03\times 10^{-7}$ & 0.96833 & $5.3\times 10^{-4}$ & 0.05122 & $2.6\times 10^{-3}$\tabularnewline
\hline 
5 & 12 & $-5.65\times 10^{-7}$ & $-3.48\times 10^{-7}$ & 0.96843 & $3.7\times 10^{-4}$ & 0.05278 & $1.8\times 10^{-3}$\tabularnewline
\hline 
5 & 13 & $-5.65\times 10^{-7}$ & $-3.99\times 10^{-7}$ & 0.96857 & $2.9\times 10^{-4}$ & 0.05388 & $1.5\times 10^{-3}$\tabularnewline
\hline 
5 & 14 & $-6.05\times 10^{-7}$ & $-4.59\times 10^{-7}$ & 0.96871 & $2.0\times 10^{-4}$ & 0.05476 & $1.1\times 10^{-3}$\tabularnewline
\hline 
5 & 15 & $-6.05\times 10^{-7}$ & $-5.65\times 10^{-7}$ & 0.96896 & $8.0\times 10^{-5}$ & 0.05489 & $4.3\times 10^{-4}$\tabularnewline
\hline 
5 & 16 & $-6.49\times 10^{-7}$ & $-6.49\times 10^{-7}$ & 0.96906 & $0$ & 0.05563 & $0$\tabularnewline
\hline 
6 & 0.5 & $-2.55\times 10^{-5}$ & $-1.00\times 10^{-7}$ & 0.96452 & $2.2\times 10^{-3}$ & 0.01126 & $6.1\times 10^{-3}$\tabularnewline
\hline 
6 & 1 & $-3.66\times 10^{-6}$ & $-1.15\times 10^{-7}$ & 0.96511 & $2.7\times 10^{-3}$ & 0.01198 & $4.6\times 10^{-3}$\tabularnewline
\hline 
6 & 2 & $-6.05\times 10^{-7}$ & $-1.00\times 10^{-7}$ & 0.96678 & $1.4\times 10^{-3}$ & 0.01607 & $2.6\times 10^{-3}$\tabularnewline
\hline 
6 & 3 & $-2.46\times 10^{-7}$ & $-1.00\times 10^{-7}$ & 0.96798 & $6.7\times 10^{-4}$ & 0.01916 & $1.4\times 10^{-3}$\tabularnewline
\hline 
6 & 4 & $-1.32\times 10^{-7}$ & $-1.00\times 10^{-7}$ & 0.96870 & $2.4\times 10^{-4}$ & 0.02196 & $5.8\times 10^{-4}$\tabularnewline
\hline 
\end{tabular}

\caption{Same as Table \ref{tab:Branch-1}, but for Branch 2.}
\label{tab:Branch-2}
\end{table}

The numerical results for Branch 3 are consistently the observationally
allowed region for $n=-2$. This result is confirmed by the data presented
in Table \ref{tab:Branch-3}.

\begin{table}
\begin{tabular}{|c|c|c|c|c|c|c|c|}
\hline 
$n$ & $m$ & $\theta_{\mathrm{min}}$ & $\theta_{\mathrm{max}}$ & $n_{s}$ & $\sigma_{n_{s}}$ & $r$ & $\sigma_{r}$\tabularnewline
\hline 
\hline 
-2 & -16 & $1.25\times 10^{0}$ & $9.33\times 10^{0}$ & 0.96309 & $1.9\times 10^{-3}$ & 0.00720 & $4.0\times 10^{-3}$\tabularnewline
\hline 
-2 & -15 & $7.71\times 10^{-1}$ & $8.12\times 10^{0}$ & 0.96379 & $2.0\times 10^{-3}$ & 0.00550 & $3.7\times 10^{-3}$\tabularnewline
\hline 
-2 & -14 & $9.49\times 10^{-1}$ & $5.75\times 10^{0}$ & 0.96399 & $1.8\times 10^{-3}$ & 0.00482 & $2.2\times 10^{-3}$\tabularnewline
\hline 
-2 & -13 & $7.20\times 10^{-1}$ & $5.00\times 10^{0}$ & 0.96461 & $1.7\times 10^{-3}$ & 0.00392 & $2.1\times 10^{-3}$\tabularnewline
\hline 
-2 & -12 & $9.49\times 10^{-1}$ & $5.00\times 10^{0}$ & 0.96472 & $1.4\times 10^{-3}$ & 0.00357 & $2.0\times 10^{-3}$\tabularnewline
\hline 
-2 & -11 & $8.86\times 10^{-1}$ & $4.06\times 10^{0}$ & 0.96485 & $1.3\times 10^{-3}$ & 0.00300 & $1.3\times 10^{-3}$\tabularnewline
\hline 
-2 & -10 & $3.36\times 10^{-1}$ & $3.54\times 10^{0}$ & 0.96430 & $1.1\times 10^{-3}$ & 0.00303 & $1.3\times 10^{-3}$\tabularnewline
\hline 
-2 & -9 & $1.44\times 10^{0}$ & $2.88\times 10^{0}$ & 0.96435 & $6.7\times 10^{-4}$ & 0.00292 & $6.7\times 10^{-4}$\tabularnewline
\hline 
-2 & -8 & $5.85\times 10^{-1}$ & $2.34\times 10^{0}$ & 0.96471 & $7.7\times 10^{-4}$ & 0.00218 & $6.4\times 10^{-4}$\tabularnewline
\hline 
-2 & -7 & $3.36\times 10^{-1}$ & $1.65\times 10^{0}$ & 0.96579 & $4.5\times 10^{-4}$ & 0.00142 & $6.1\times 10^{-4}$\tabularnewline
\hline 
-2 & -6 & $2.73\times 10^{-1}$ & $1.25\times 10^{0}$ & 0.96547 & $6.4\times 10^{-4}$ & 0.00107 & $4.4\times 10^{-4}$\tabularnewline
\hline 
-2 & -5 & $2.11\times 10^{-2}$ & $8.86\times 10^{-1}$ & 0.96592 & $5.5\times 10^{-4}$ & 0.00068 & $3.4\times 10^{-4}$\tabularnewline
\hline 
-2 & -4 & $3.60\times 10^{-1}$ & $3.60\times 10^{-1}$ & 0.96624 & $0$ & 0.00041 & $0$\tabularnewline
\hline 
-2 & -3 & $2.07\times 10^{-1}$ & $2.73\times 10^{-1}$ & 0.96626 & $8.0\times 10^{-5}$ & 0.00026 & $4.0\times 10^{-5}$\tabularnewline
\hline 
-2 & -0.5 & $4.59\times 10^{-3}$ & $4.59\times 10^{-3}$ & 0.96661 & $0$ & 0.00001 & $0$\tabularnewline
\hline 
-2 & 0.5 & $2.98\times 10^{-2}$ & $4.43\times 10^{-1}$ & 0.96819 & $7.5\times 10^{-4}$ & 0.00023 & $1.4\times 10^{-4}$\tabularnewline
\hline 
-2 & 1 & $1.57\times 10^{-1}$ & $9.49\times 10^{-1}$ & 0.96822 & $6.8\times 10^{-4}$ & 0.00048 & $2.7\times 10^{-4}$\tabularnewline
\hline 
-2 & 2 & $1.37\times 10^{-1}$ & $1.90\times 10^{0}$ & 0.96821 & $7.6\times 10^{-4}$ & 0.00101 & $6.1\times 10^{-4}$\tabularnewline
\hline 
-2 & 3 & $1.37\times 10^{-1}$ & $3.08\times 10^{0}$ & 0.96795 & $7.4\times 10^{-4}$ & 0.00124 & $9.4\times 10^{-4}$\tabularnewline
\hline 
-2 & 4 & $2.73\times 10^{-1}$ & $4.67\times 10^{0}$ & 0.96831 & $7.2\times 10^{-4}$ & 0.00240 & $1.3\times 10^{-3}$\tabularnewline
\hline 
-2 & 5 & $4.43\times 10^{-1}$ & $6.16\times 10^{0}$ & 0.96816 & $6.5\times 10^{-4}$ & 0.00271 & $1.6\times 10^{-3}$\tabularnewline
\hline 
-2 & 6 & $6.27\times 10^{-1}$ & $8.12\times 10^{0}$ & 0.96845 & $4.9\times 10^{-4}$ & 0.00437 & $1.7\times 10^{-3}$\tabularnewline
\hline 
-2 & 7 & $8.27\times 10^{-1}$ & $1.07\times 10^{1}$ & 0.96837 & $5.5\times 10^{-4}$ & 0.00517 & $2.4\times 10^{-3}$\tabularnewline
\hline 
-2 & 8 & $4.14\times 10^{-1}$ & $1.52\times 10^{1}$ & 0.96839 & $5.8\times 10^{-4}$ & 0.00663 & $3.1\times 10^{-3}$\tabularnewline
\hline 
-2 & 9 & $1.44\times 10^{0}$ & $2.14\times 10^{1}$ & 0.96851 & $4.2\times 10^{-4}$ & 0.00914 & $3.7\times 10^{-3}$\tabularnewline
\hline 
-2 & 10 & $1.17\times 10^{0}$ & $1.49\times 10^{2}$ & 0.96870 & $4.4\times 10^{-4}$ & 0.02394 & $1.5\times 10^{-2}$\tabularnewline
\hline 
-2 & 11 & $7.71\times 10^{-1}$ & $1.39\times 10^{2}$ & 0.96830 & $1.0\times 10^{-3}$ & 0.02167 & $1.5\times 10^{-2}$\tabularnewline
\hline 
-2 & 12 & $7.20\times 10^{-1}$ & $1.30\times 10^{2}$ & 0.96815 & $3.8\times 10^{-4}$ & 0.02310 & $1.6\times 10^{-2}$\tabularnewline
\hline 
-2 & 13 & $8.27\times 10^{-1}$ & $1.21\times 10^{2}$ & 0.96791 & $3.8\times 10^{-4}$ & 0.02250 & $1.7\times 10^{-2}$\tabularnewline
\hline 
-2 & 14 & $1.09\times 10^{0}$ & $1.13\times 10^{2}$ & 0.96781 & $3.2\times 10^{-4}$ & 0.02469 & $1.6\times 10^{-2}$\tabularnewline
\hline 
-2 & 15 & $6.27\times 10^{-1}$ & $1.05\times 10^{2}$ & 0.96770 & $3.2\times 10^{-4}$ & 0.02441 & $1.4\times 10^{-2}$\tabularnewline
\hline 
-2 & 16 & $1.09\times 10^{0}$ & $1.05\times 10^{2}$ & 0.96750 & $3.5\times 10^{-4}$ & 0.02457 & $1.6\times 10^{-2}$\tabularnewline
\hline 
-3 & -16 & $3.79\times 10^{0}$ & $3.73\times 10^{1}$ & 0.96775 & $9.2\times 10^{-4}$ & 0.01680 & $4.9\times 10^{-3}$\tabularnewline
\hline 
-3 & -15 & $4.36\times 10^{0}$ & $3.25\times 10^{1}$ & 0.96787 & $7.3\times 10^{-4}$ & 0.01604 & $4.2\times 10^{-3}$\tabularnewline
\hline 
-3 & -14 & $5.00\times 10^{0}$ & $2.64\times 10^{1}$ & 0.96811 & $5.5\times 10^{-4}$ & 0.01483 & $3.3\times 10^{-3}$\tabularnewline
\hline 
-3 & -13 & $5.75\times 10^{0}$ & $2.00\times 10^{1}$ & 0.96838 & $4.0\times 10^{-4}$ & 0.01342 & $2.4\times 10^{-3}$\tabularnewline
\hline 
-3 & -12 & $6.16\times 10^{0}$ & $1.62\times 10^{1}$ & 0.96865 & $2.5\times 10^{-4}$ & 0.01237 & $1.7\times 10^{-3}$\tabularnewline
\hline 
-3 & -11 & $7.58\times 10^{0}$ & $1.23\times 10^{1}$ & 0.96890 & $7.0\times 10^{-5}$ & 0.01143 & $7.9\times 10^{-4}$\tabularnewline
\hline 
-4 & -16 & $6.95\times 10^{1}$ & $2.26\times 10^{2}$ & 0.96855 & $2.8\times 10^{-4}$ & 0.02101 & $2.2\times 10^{-3}$\tabularnewline
\hline 
-4 & -15 & $8.56\times 10^{1}$ & $1.71\times 10^{2}$ & 0.96884 & $1.1\times 10^{-4}$ & 0.01972 & $1.3\times 10^{-3}$\tabularnewline
\hline 
\end{tabular}

\caption{Same as Table \ref{tab:Branch-1}, but for Branch 3.}
\label{tab:Branch-3}
\end{table}

We find that the results for $n=-2$ show excellent agreement with
those obtained from the approximate analytical treatment(e.g., $n_{s}=0.966837,r\simeq0.00109071$
for $\theta\sim1$) . 

A comprehensive summary of the results is provided in Table \ref{tab:Total}.
Branch 1, Branch 2 and Branch 3 all yield solutions that are consistent
with observational bounds over a wide range of $n$ and $m$, and
in particular, $r$. 

It is also observed that larger values of $n$ tend to satisfy the
observational bounds with smaller values of $\theta$, and that smaller
values of $m$ are effective in realizing lower values of $r$.

\begin{table}
\begin{tabular}{|c|c|c|}
\hline 
 & $n_{s}$ & $r$\tabularnewline
\hline 
\hline 
branch1 & $0.96743\pm0.00136$ & $0.03819\pm0.01298$\tabularnewline
\hline 
branch2 & $0.96655\pm0.00208$ & $0.03193\pm0.01605$\tabularnewline
\hline 
branch3 & $0.96754\pm0.00162$ & $0.01492\pm0.01386$\tabularnewline
\hline 
\end{tabular}

\caption{Summary of the predicted spectral index $(n_{s})$ and tensor-to-scalar
ratio $(r)$ for each branch, together with their uncertainties. \label{tab:Total}}
\end{table}

\begin{figure}
\centering
\includegraphics[width=0.5\textwidth,totalheight=1\textheight,keepaspectratio]{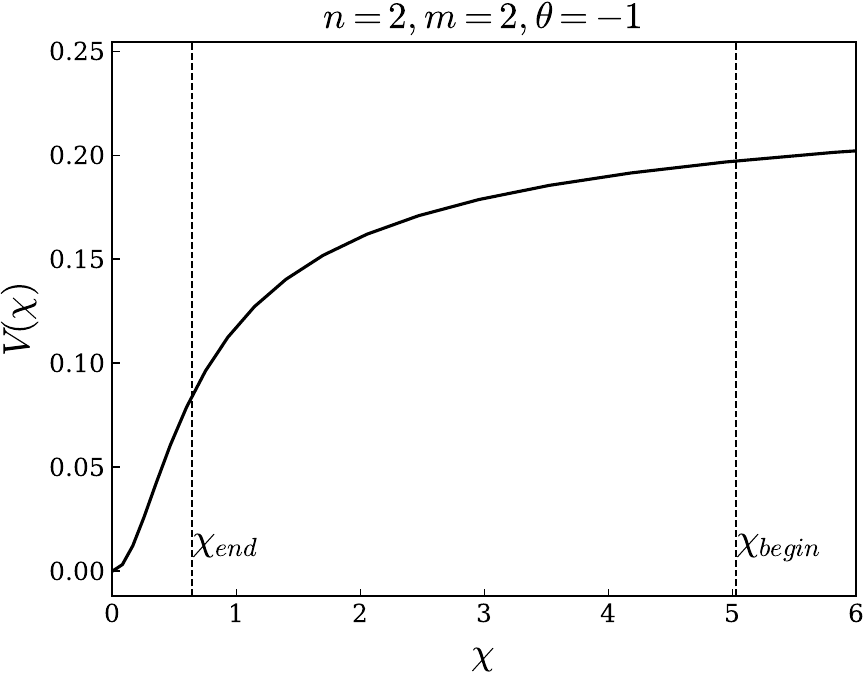}

\caption{\textcolor{black}{Potential $V(\chi)$ for $n=2$, $m=2$, and $\theta=-1$. 
The horizontal axis represents the canonically normalized inflaton field $\chi$, 
and the vertical axis shows the potential $V(\chi)$. 
The vertical dashed lines indicate the horizon-exit point and the end point of the 
slow-roll phase. 
The potential is smooth in the relevant slow-roll interval and does not show 
rapid non-monotonic undulations in this region.\label{fig:Potentialn2}}}
\end{figure}

\textcolor{black}{Figure \ref{fig:Potentialn2} shows the potential for $n = 2$, $m = 2$, and $\theta = -1$, along with the starting and ending points of the slow-roll phase and the evolution of the inflaton field.
We also inspected whether the Virasoro-transformed potential develops non-monotonic
undulations or inflection-point-like features. 
For the representative parameter set shown in Fig.~\ref{fig:Potentialn2}, 
the potential is smooth and monotonic in the relevant slow-roll interval between the 
horizon-exit point and the end point of inflation. 
No rapid non-monotonic undulation is observed in this interval, and hence the slow-roll 
approximation remains justified for this example. 
Depending on the branch of the Virasoro transformation and on the choice of 
$(n,m,\theta)$, changes in curvature or inflection-point-like behavior may appear 
outside the relevant slow-roll region. 
A systematic classification of such features is left for future work.
}

\begin{figure}
\includegraphics[width=1\textwidth,totalheight=0.5\textheight]{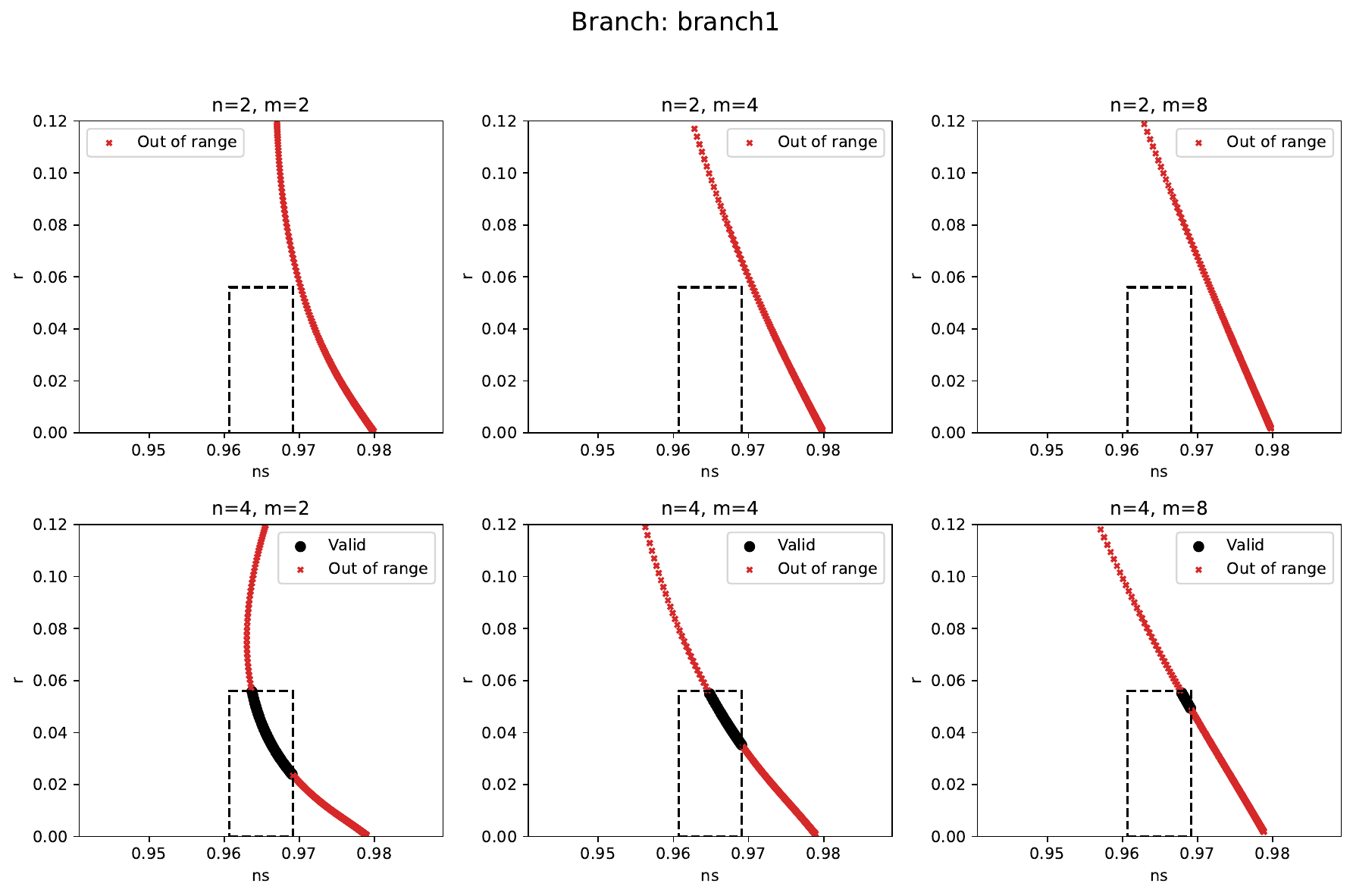}

\caption{Parametric plots of the spectral index $n_{s}$ and tensor-to-scalar
ratio $r$ for various $(n,\ m)$ in Branch 1. The solid circles indicate
parameter sets that satisfy observational constraints, while crosses
correspond to values outside the allowed region.\label{fig:Branch1Graph}}
\end{figure}

\begin{figure}
\includegraphics[width=1\textwidth,totalheight=0.5\textheight]{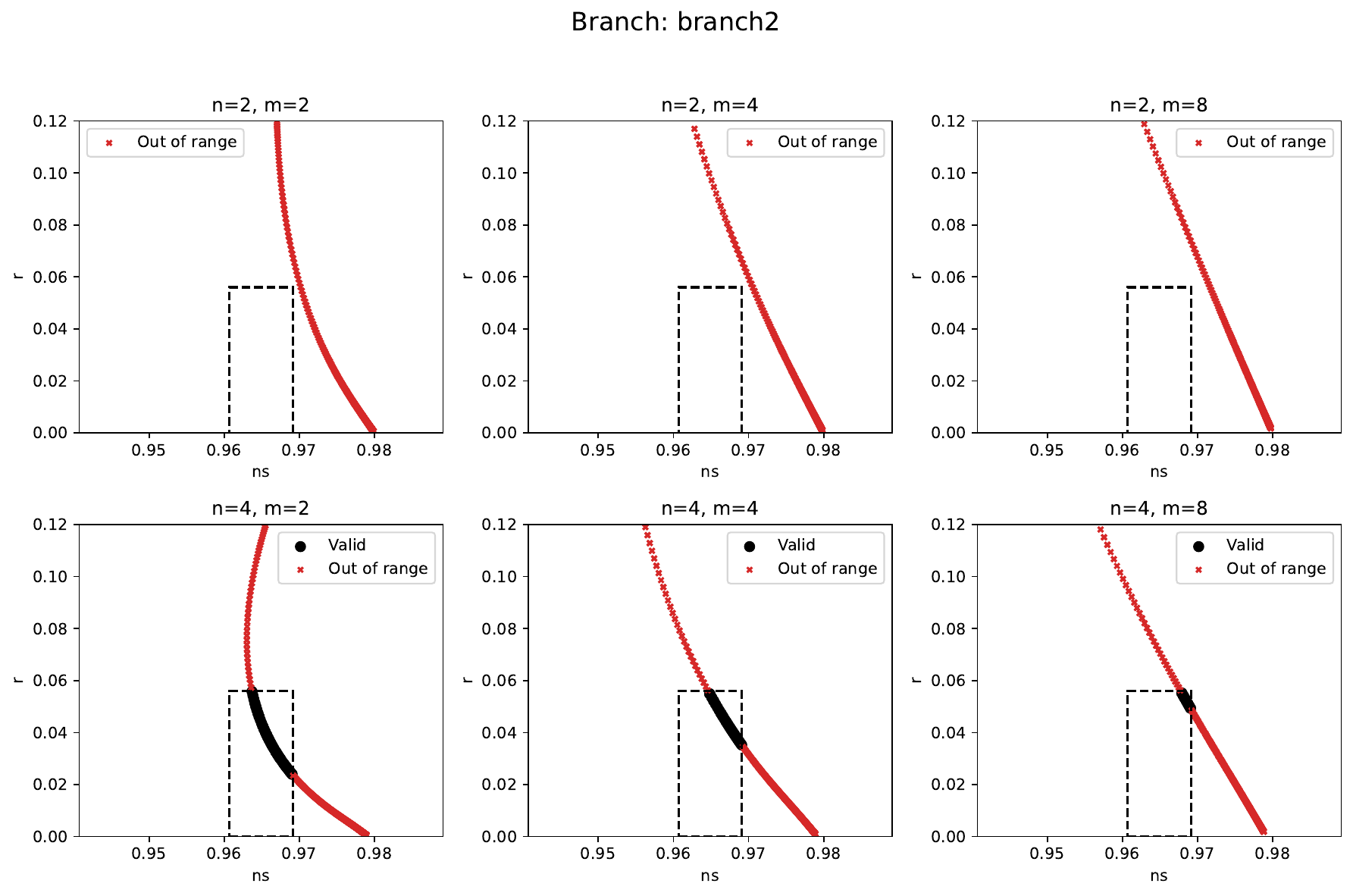}

\caption{Same as Figure \ref{fig:Branch1Graph}, but for Branch 2.\label{fig:Branch2-Graph}}
\end{figure}

Figures \ref{fig:Branch1Graph} and \ref{fig:Branch2-Graph} show
parametric plots in the $(n_{s},r)$ plane for various combinations
$(n,m)$ based on the numerical data described above. In these plots,
the parameter $\theta$ is varied, and the predicted values trace
a trajectory from high $n_{s}$ and low $r$ to lower $n_{s}$ and
higher $r$ within the region allowed by current observations.

It should be noted, however, that as $\theta$ changes, some parameter
combinations may leave the allowed region, resulting in predictions
that fall outside the observational constraints.


\begin{figure}
\includegraphics[width=1\textwidth,totalheight=0.5\textheight]{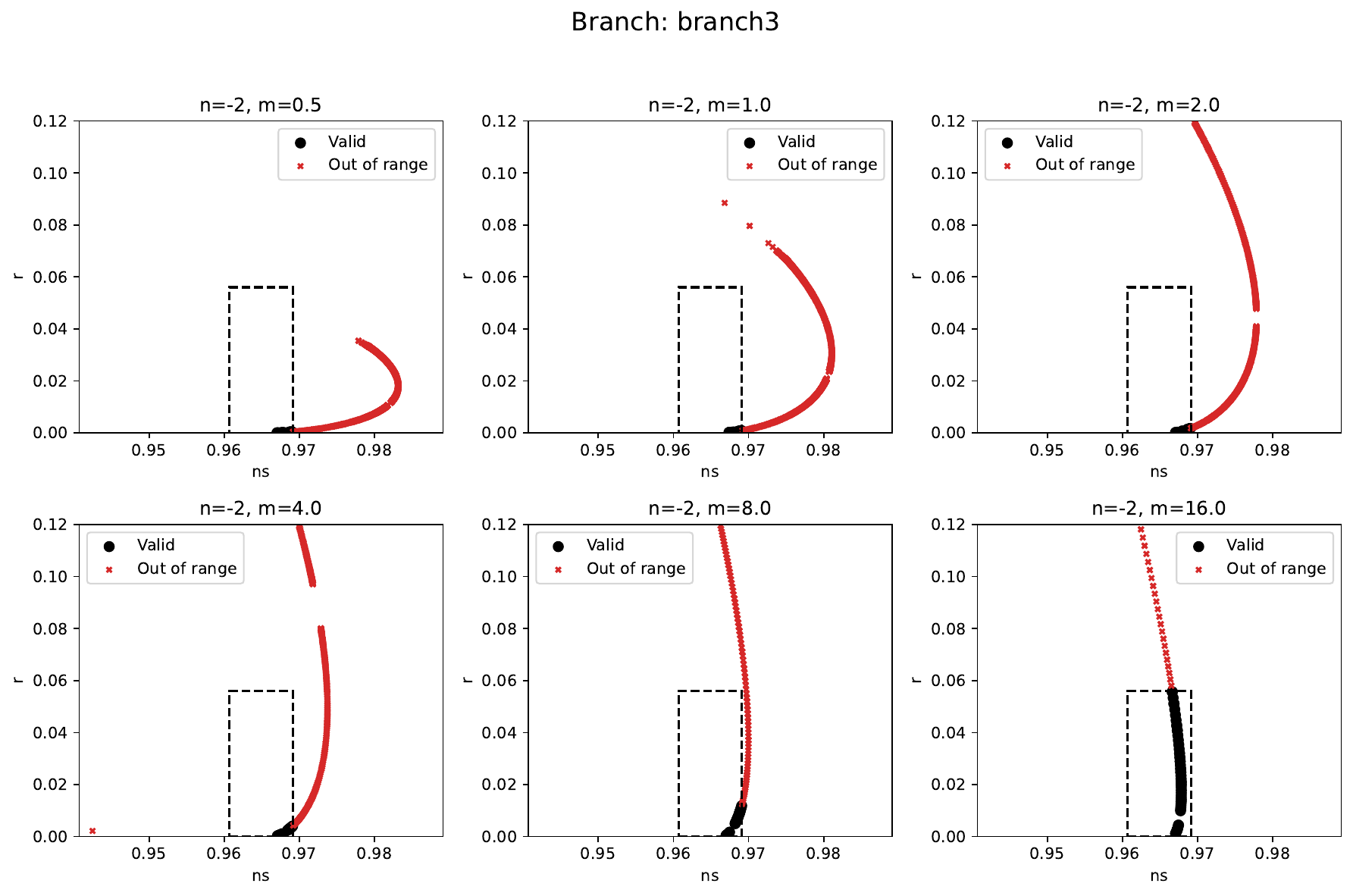}

\caption{Same as Figure \ref{fig:Branch1Graph}, but for Branch 3.\label{fig:Branch3-Graph-1}}
\end{figure}

\color{black}
Although not included in this paper, when $N_e$ is varied, the parametric plot of $(n_s, r)$ with $\theta$ as a parameter shifts in the direction of increasing $n_s$ as $N_e$ increases.
It shows a similar trend to the result obtained in the approximate analysis for the case of $n = -2$ discussed in Section 3.
\color{black}

In summary, our numerical analysis demonstrates that Branch 1, Branch
2 and Branch 3 can all realize slow-roll inflation with predictions
for $(n_{s},\ r)$ consistent with current observational constraints
over a broad range of model parameters. Notably, the ability to achieve
small values of $r$ without fine-tuning is a distinctive feature
of this framework. At the same time, the results highlight the sensitivity
of the predictions to the choice of the Virasoro index $n$, the potential
power $m$, and the squeezing parameter $\theta$, with some parameter
combinations falling outside the observationally allowed region. In
particular, the value of $r\lesssim0.03$ achieved in Branch 3 is
noteworthy. This value is not excluded by the results of Planck PR4
+ BK18.

In order to perform more accurate calculations beyond the branched
point, it is necessary to consider inflation as a complex scalar field.
In this case, contributions from the curvature of the inflaton beyond
the horizon also come into play, and complex corrections are added
to the slow-roll parameter. These accurate calculations will improve
the predictive power of the theory\cite{Wands2002}.

These findings suggest that Virasoro-squeezed inflation models provide
a flexible and unifying framework for inflationary phenomenology.
Further studies, including more systematic exploration of the parameter
space and direct comparison with upcoming cosmological observations,
would be valuable for assessing the full predictive power and testability
of the model.

\section{Overview of Approximate Analytical Results}

In previous calculations, for the cases $n\neq-2$, the potential
can be approximated as
\begin{eqnarray}
V_{\chi}^{(m,n)}(\chi_{n})\approx\frac{\lambda_{m}}{m}\left|\frac{A_{n}\chi_{n}^{\frac{1}{n+2}}}{\left(1-n\theta(A_{n})^{n}\chi_{n}^{\frac{n}{n+2}}\right)^{1/n}}\right|^{m}.\label{n>0-1}
\end{eqnarray}
This approximation allows us to analyze, in a parametric way, how
the predictions for $n_{s}$ and $r$ depend on the parameters $n,m,\theta$
and the e-folding number $N_{e}$.

The details of the derivation are provided in Appendix B\footnote{Here, $M_{pl}\equiv1/\sqrt{8\pi G}$ denotes the reduced Planck mass.}.
Here, we summarize the main results. Using the standard slow-roll
inflationary formalism, the spectral index $n_{s}$ and the tensor-to-scalar
ratio $r$ can be approximated as 
\begin{equation}
n_{s}=1-M_{pl}^{2}\left(\frac{m}{n+2}\frac{1}{\chi_{n}}-\frac{m}{n}\delta_{1}\right)^{2}+2M_{pl}^{2}\left(-\frac{m}{n+2}\frac{1}{\chi_{n}}-\frac{m}{n}\delta_{2}\right),\label{eq:nsr}
\end{equation}

\begin{equation}
r=8M_{pl}^{2}\left(\frac{m}{n+2}\frac{1}{\chi_{n}}-\frac{m}{n}\delta_{1}\right)^{2}.\label{eq:nsr2}
\end{equation}
Here, $\delta_{1}$ and $\delta_{2}$ collect the higher-order corrections
arising from the derivatives of $V(\chi_{n})$; see Appendix B for
full details. Explicitly, they are given by

\begin{equation}
\delta_{1}\equiv\frac{-n\theta A_{n}^{n}\frac{n}{n+2}\chi_{n}^{\frac{-2}{n+2}}}{1-n\theta A_{n}^{n}\chi_{n}^{\frac{n}{n+2}}},
\end{equation}

\begin{equation}
\delta_{2}\equiv\frac{-n\theta A_{n}^{n}\frac{-2n}{\left(n+2\right)^{2}}\chi_{n}^{\frac{-2}{n+2}-1}}{1-n\theta A_{n}^{n}\chi^{\frac{n}{n+2}}}+\frac{-\frac{n^{3}}{(n+2)^{2}}\theta^{2}A_{n}^{2n}\chi_{n}^{\frac{-4}{n+2}}}{\left(1-n\theta A_{n}^{n}\chi_{n}^{\frac{n}{n+2}}\right)^{2}}.
\end{equation}
The variable $\chi_{n}$ represents the value of the inflaton field
at the beginning of inflation, which can be determined from the e-folding
number $N_{e}$ and the field value $\chi_{end}$ at the end of inflation
(defined by $\epsilon_{\mathrm{end}}=1$). The behaviors of $N,\chi$
and $\chi_{end}$ differ for the case $-2<n<0$ compared to other
ranges; see Appendix B for further discussion.

Our numerical analysis in the previous section revealed that solutions
consistent with observations are found for positive $n$ with negative
$\theta$, while for negative $n$ with positive $\theta$, no viable
solutions within the observational bounds were obtained in the scanned
range. This behavior can also be qualitatively understood from the
approximate analytical calculation: the asymptotic structure of the
potential and the dependence of $n_{s}$ and $r$ on $(n,\theta)$
follow similar trends.

\begin{figure}
\centering
\includegraphics[width=0.8\textwidth,totalheight=0.8\textwidth,keepaspectratio]{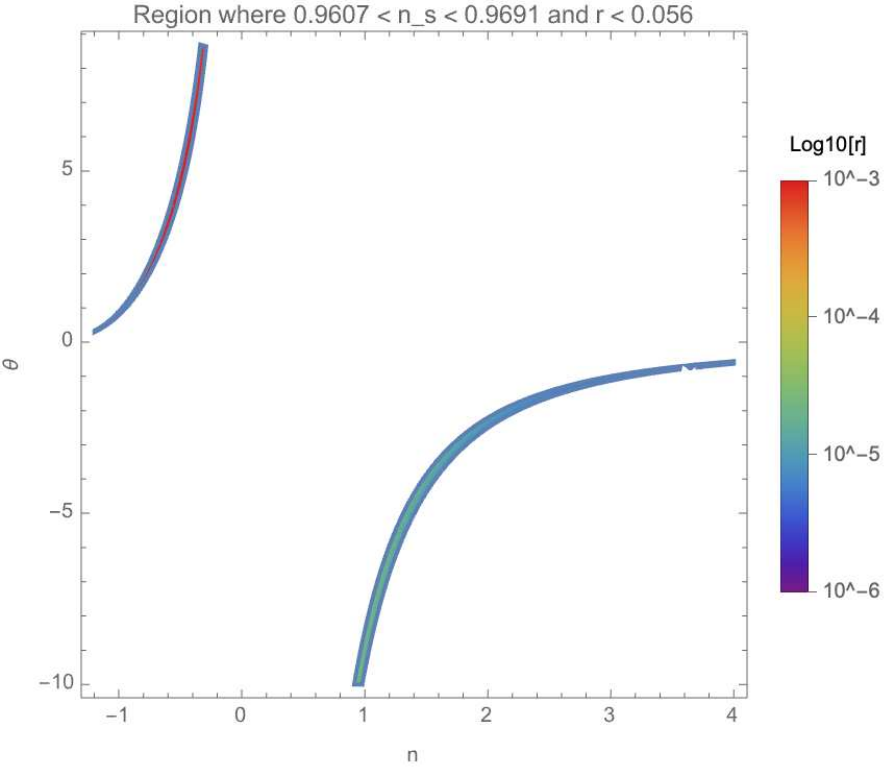}

\caption{Allowed region in the $(n,\text{\ensuremath{\theta}})$ parameter
space where the predicted spectral index $n_{s}$ and tensor-to-scalar
ratio $r$ satisfy the observational constraints $0.9607<n_{s}<0.9691$
and $r<0.056$. The color scale represents $\log_{10}r$.\label{fig:nthetaplane}}
\end{figure}

Figure \ref{fig:nthetaplane} displays the allowed region in the $(n,\theta)$-plane
where the predicted values of $(n_{s},\ r)$ satisfy $0.9607<n_{s}<0.9691$
and $r<0.056$. The color scale indicates $\log_{10}r$, with smaller
values of $r$ shown in blue and larger values in red. It is apparent
that solutions consistent with observations are only found for positive
$n$ with negative $\theta$, while for negative $n$, no viable solutions
exist within the scanned parameter range. 

From the sign of $n$, one can see that the direction of squeezing
(i.e., the action of the Virasoro generator $L_{n}$) must be reversed
for negative $n$ compared to positive $n$. As $n$ increases, the
squeezing effect becomes stronger, enabling the observational constraints
to be satisfied even for smaller values of $\theta$. However, excessive
squeezing can cause the predictions to deviate from the observed values.
Overall, for positive $n$, a broader range of $\theta$ satisfies
the observational constraints, and the resulting scalar-to-tensor
ratio $r$ can be extremely small.

\section{Conclusion and Discussion}

In this paper, we have proposed a new class of inflationary models,
termed ``Virasoro inflation,'' constructed by applying a conformal
transformation to a complex scalar field. The transformation is generated
by Virasoro generators $L_{n}$ $(n=0,\pm1,\pm2,\cdots)$ and parametrized
by a continuous parameter $\theta$. so that each model is specified
by the pair $(n,\theta)$.

A key assumption in our construction is that the conformal transformation
is canonical, preserving the canonical conjugate relations. This naturally
induces squeezing-like effects in the scalar field sector.

Under this framework, the original scalar potential $V(\phi^{\dagger}\phi)=\frac{\lambda_{m}}{m}|\phi(t)|^{m}$
is mapped into a new potential $V_{\chi}^{(m)}(\chi_{n})$ for a redefined
field $\chi_{n}(t)$, which ensures the kinetic term remains canonical.
The modulus of the new field, $|\chi_{n}(t)|$ plays the role of the
inflaton, slowly rolling down the transformed potential $V_{\chi}^{(m)}(\chi_{n})$.

This construction yields a wide variety of novel inflationary models.
Among them, models with positive $n$ are shown to produce predictions
for the spectral index $n_{s}$ and tensor-to-scalar ratio $r$ that
are consistent with current CMB observations, under suitable choices
of $(m,\ \theta)$ and for typical e-folding numbers ($N_{e}=50\mathrm{-}60$).
In this paper, we have presented detailed results for $N_{e}=60$
as a representative case.

A notable advantage of our framework is its simplicity: the transformation
acts solely on the scalar field, without any need for conformal coupling
to gravity. Moreover, the plateau structure of the transformed potential
emerges generically from the maximum modulus theorem in complex analysis.
Under a holomorphic and singularity-free transformation 

\begin{equation}
\phi\to\chi=F(\phi),
\end{equation}
the potential 
\begin{equation}
V_{\chi}(\chi)=|W(\chi)|^{2}=V_{\phi}(F(\phi)^{\dagger}F(\phi))
\end{equation}
attains its maximum only on the boundary $|\chi|\to\infty$. Therefore,
the interior of the field space exhibits a natural plateau.

We have also assumed that the squeezing effect occurs just before
the onset of inflation. As a possible physical origin, we briefly
discussed analogies with higher-order squeezing mechanisms in quantum
optics.

It is intriguing to speculate whether the conformal squeezing studied
here might be related to effective dynamics in brane systems, such
as M2--M5 configurations, where Virasoro-like symmetries and holomorphic
deformations naturally emerge. One possible manifestation of such
effective dynamics is the appearance of friction-like terms in the
inflaton's evolution. These terms do not correspond to conventional
dissipation, but may instead encode geometric resistance arising from
spontaneous symmetry breaking or nontrivial twisting between intersecting
branes, leading to effective non-conservative behavior at low energies.

We leave a detailed investigation of these possible connections, as
well as further phenomenological implications of Virasoro inflation,
as interesting directions for future research.

\section*{Acknowledgments}

The authors would like to thank Shiro Komata for carefully reading
the manuscript and providing valuable comments. They are also grateful
to the members of the theoretical seminar group at the Open University
of Japan for their helpful feedback and for providing a stimulating
environment in which to share ongoing progress.

\appendix

\section{Numerical Setup Detail}

\subsection{Numerical Procedure}

To evaluate the inflationary observables $n_{s}$ and $r$ for a given
Virasoro inflation model, we follow a numerical procedure that avoids
the analytical approximations used in Section $4$. The key idea is
to solve the end-of-inflation condition and the e-folding number integral
numerically, based on the transformed potential.

The overall algorithm proceeds as follows:\noindent
\vspace{1.0em} 
\hrule
\begin{tcolorbox}[highlight algorithm]  
\textbf{Algorithm: Numerical Procedure for Computing \( n_s \) and \( r \)}

\begin{algorithmic}[1]
\State \textbf{Input:} Virasoro index \( n \), potential power \( m \), dimensionless parameter \( \bar{\theta} \), e-folding number \( N_e \)
\State Select branch depending on sign of \( n \):
  \If{\( n > 0 \)}
    \State Use Branch 2+
  \Else
    \State Use Branch 3$-$
  \EndIf
\State Define dimensionless variables \( y = \bar{F}_n(\phi) \), \( x = \bar{\chi}_n \)
\State Obtain analytical forms of \( y(x) \), \( y'(x) \), and \( y''(x) \) from Appendix~B
\State Solve for \( y_e \) satisfying:
  \[
    \epsilon_V(y_e) = \frac{1}{2} m^2 |n|^{2/n} |\bar{\theta}|^{2/n} \left( \frac{y'}{y} \right)^2 = 1
  \]
\State Solve for \( y_b \) using the e-folding condition:
  \[
    N_e = \frac{1}{m |n \bar{\theta}|^{2/n}} \int_{y_e}^{y_b} \frac{y}{(y')^2} \, dy
  \]
\State Compute slow-roll parameter at \( y = y_b \):
  \[
    \epsilon_V = \frac{1}{2} m^2 |n|^{2/n} |\bar{\theta}|^{2/n} \left( \frac{y'}{y} \right)^2
  \]
  \[
    \eta_V = |n|^{2/n} |\bar{\theta}|^{2/n} \left[ m(m - 1) \left( \frac{y'}{y} \right)^2 + m \frac{y''}{y} \right]
  \]
\State Evaluate observables:
  \[
    n_s = 1 - 6\epsilon_V + 2\eta_V, \qquad r = 16\epsilon_V
  \]
\State \textbf{Output:} \( n_s \), \( r \)
\end{algorithmic}

\end{tcolorbox}
\vspace{0.5em}\noindent\hrule
\vspace{1.0em}

This procedure enables us to explore the full parameter space $(n,m,\bar{\theta})$,
including the subtleties arising from branch selection in the Virasoro
transformation. Analytical forms for $y,\ y'$ and $y''$ specific
to each branch are listed in next Section.

\subsection{List of $y$}

The conformal transformation $\phi\to F_{n}(\phi;\theta)$ is generated
by the $n$-th Virasoro generator with a complex parameter $\theta$,
and is given by
\begin{equation}
F_{n}(\phi;\theta)=\phi_{n}(\theta)=\frac{\phi}{(1-n\theta\phi^{n})^{1/n}}.
\end{equation}

This conformal transformation maps the original complex scalar field
model defined by $\phi$ into a new scalar field model characterized
by $\chi_{n}$, which plays the role of the inflaton field in the
transformed theory.

The relation between $\phi$ and $\chi$ is given by
\begin{equation}
\chi_{n}(\phi)=\int_{0}^{\phi}\frac{d\phi}{F'_{n}(\phi)}=\int_{0}^{\phi}d\phi(1-n\theta\phi^{n})^{1+1/n}=\int_{0}^{\phi}d\phi\frac{\phi^{n+1}}{F_{n}(\phi)^{n+1}}.
\end{equation}

The potential $V_{\chi}$ is derived from the original potential $V_{\phi}=\frac{\lambda}{m}|\phi|^{m}$
as

\begin{equation}
V_{\chi}=\frac{\lambda}{m}|F_{n}(\phi)|^{m}.
\end{equation}

All fields are assumed to have the same mass dimension $[M]$, while
the potential has mass dimension $[M^{4}]$.

The mass dimension of the scalar field in four-dimensional spacetime
is $[\phi]=[M]$. From the form of the conformal transformation, the
mass dimension of $\theta$ is given by
\begin{equation}
[\theta]=[M^{-n}].
\end{equation}

The field $\chi$ has the same dimension as $\phi$, namely $[\chi]=[M]$.

We now define the corresponding dimensionless quantities $\bar{\phi}$,
$\bar{\chi}_{n}$, $\bar{F}_{n}$ and $\bar{V}_{\chi}$ by absorbing
the common factor $n\theta$ as follows:
\begin{equation}
|n\theta|\phi^{n}=\bar{\phi}^{n},\ \mathrm{or}\ \mathrm{equivalently}\ \phi=\frac{\bar{\phi}}{|n\theta|^{1/n}},
\end{equation}

\begin{equation}
\chi=\frac{\bar{\chi}}{|n\theta|^{1/n}},\ F_{n}=\frac{\bar{F_{n}}}{|n\theta|^{1/n}},\ V_{\chi}=\frac{\bar{V_{\chi}}}{|n\theta|^{4/n}}.
\end{equation}

Here, we introduce a dimensionless squeezing parameter $\bar{\theta}=\theta M_{pl}^{n}$.
Thus, the model is fully specified by the set of parameters $(n,m,\bar{\theta})$.
We express all relevant equations in terms of two dimensionless variables,
$\bar{F}_{n}=y\ \mathrm{and}\ \bar{\chi}_{n}=x$, so that the formulation
can be consistently described using the pair $(x,y)$. When performing
the integration, we change the integration variable from $\phi$ to
a new variable $t$.

The transformation rule for the integration variable is given as follows:
\begin{itemize}
\item For $n>0$: Since $\phi$ increases monotonically,
\begin{equation}
t=|n\theta|\phi^{n}\equiv\bar{\phi}^{n}.
\end{equation}
\item For $n<0$: As $\phi$ becomes large, $\phi^{|n|}$ becomes small,
and therefore
\begin{equation}
\frac{1}{t}=|n\theta|\phi^{-|n|}\ \mathrm{or}\ t=\frac{\phi^{|n|}}{|n\theta|}=\bar{\phi}^{|n|}.
\end{equation}
\end{itemize}
The conformal transformation
\begin{equation}
F_{n}(\phi;\theta)=\phi_{n}(\theta)=\frac{\phi}{(1-n\theta\phi^{n})^{1/n}}
\end{equation}
can be rewritten using the rescaled variables introduced earlier.
Specifically, we define
\begin{equation}
y=\bar{F}_{n}=|n\theta|^{1/n}F_{n}(\phi;\theta)=|n\theta|^{1/n}\phi_{n}(\theta)=\frac{\bar{\phi}}{(1-\mathrm{sgn(n\theta)}\bar{\phi}^{n})^{1/n}}.
\end{equation}

Here, we introduce $\epsilon=\mathrm{sgn}(\theta)$.

Thus, we have
\begin{equation}
y=\frac{\bar{\phi}}{(1-\mathrm{sgn}(n\theta)\bar{\phi}^{n})^{1/n}}.
\end{equation}

We now re-express this equation in terms of the variable $t$.

At this point, the sign of $n$ introduces branching behavior, since
the denominator involves an $n$-th root, which requires careful branch
selection. The choice of branch determines whether the transformed
potential exhibits a plateau structure, which is essential for satisfying
the slow-roll conditions.
\begin{itemize}
\item For $n>0$ (denoted as Branch2+), we use $t=|n\theta|\phi^{n}\equiv\bar{\phi}^{n}$.
Since $\mathrm{sgn}(n\theta)=\epsilon$, we have
\begin{equation}
y=\frac{t^{1/n}}{(1-\epsilon t)^{1/n}}=t^{1/n}(1-\epsilon t)^{1/n}.
\end{equation}
\item For $n<0$ (denoted as Branch3-), we use $t=\bar{\phi}^{|n|}$, and
since $\mathrm{sgn}(n\theta)=-\epsilon$, we obtain
\begin{equation}
y=\frac{t^{1/|n|}}{(1+\epsilon t^{-1})^{-1/|n|}}=\left(\frac{t^{-1}}{(1+\epsilon t^{-1})}\right)^{-1/|n|}=(t+\epsilon)^{1/|n|}.
\end{equation}
\item {[}Branch1+{]} uses the same substitution as Branch 2+, namely $t=|n\theta|\phi^{n}$.
However, it differs in the way the $n$-th root in the denominator
is taken. We start by using the identity
\begin{equation}
y=\frac{t^{1/n}}{(1-\epsilon\text{t})^{1/n}},
\end{equation}
which corresponds to the principal branch used in Branch 2+. Here,
we consider an alternative way of taking the $n$-th root. By rewriting
the denominator, we obtain
\begin{equation}
(1-\epsilon t)^{1/n}=(-\epsilon(-\frac{1}{\epsilon}+t))^{1/n}=(-\epsilon)^{1/n}(t-\epsilon)^{1/n},
\end{equation}
where we have used the identity $(-\epsilon)\epsilon=-1$. With this
substitution, the expression for $y$ becomes
\begin{equation}
y=\frac{t^{1/n}}{(-\epsilon)^{1/n}(t-\epsilon)^{1/n}}.
\end{equation}
 This is still consistent with the form of Branch 2+, but in Branch
1+, we take a different branch of the $n$-th root such that the prefactor
$(-\epsilon)^{-1/n}$ is eliminated. This leads to the final expression
\begin{equation}
y=\frac{t^{1/n}}{(t-\epsilon)^{1/n}}.
\end{equation}
\end{itemize}
To summarize, the three branches are defined as follows:
\begin{itemize}
\item {[}Branch1+{]} (for $n>0)$ 
\begin{equation}
\begin{aligned}\mathrm{Variable\ substitution:} & t=|n\theta|\phi^{n}\equiv\bar{\phi}^{n}\\
\mathrm{Resulting\ expression:} & y=\frac{t^{1/n}}{(t-\epsilon)^{1/n}}
\end{aligned}
\end{equation}
\item {[}Branch2+{]} (for $n>0)$
\begin{equation}
\begin{aligned}\mathrm{Variable\ substitution:} & t=|n\theta|\phi^{n}\equiv\bar{\phi}^{n}\\
\mathrm{Resulting\ expression:} & y=t^{1/n}(1-\epsilon t)^{-1/n}
\end{aligned}
\end{equation}
\item {[}Branch3-{]} (for $n<0)$
\begin{equation}
\begin{aligned}\mathrm{Variable\ substitution:} & t=\frac{\phi^{|n|}}{|n\theta|}=\bar{\phi}^{|n|}\\
\mathrm{Resulting\ expression:} & y=(t+\epsilon)^{1/|n|}
\end{aligned}
\end{equation}
\end{itemize}

\color{black}
\subsection{Construction of the statistical sample}

In this subsection, we describe how the statistical samples used to compute
$\sigma_{n_s}$ and $\sigma_r$ in Section~4.2 are constructed.

For each fixed pair $(n,m)$ and for each branch of the Virasoro transformation, we proceed as follows.

\begin{enumerate}
\item We scan the squeezing parameter $\theta$ over a prescribed range. 
The range is chosen so as to cover the values for which the transformed potential is regular and the resulting observables may enter the observationally allowed region.

\item For each value of $\theta$, we follow the numerical procedure described in Appendix~A.1. 
First, the end point of inflation \(y_e\) is determined by solving
\[
\epsilon_V(y_e)=1 .
\]

\item The field value \(y_b\) at horizon exit is then obtained by solving the e-folding condition
\[
N_e
=
\int_{y_e}^{y_b} \cdots\, dy ,
\]
where in the numerical analysis of Section~4 we set \(N_e=60\).

\item At \(y=y_b\), we evaluate the slow-roll parameters \(\epsilon_V\) and \(\eta_V\), and compute the observables
\[
n_s = 1-6\epsilon_V+2\eta_V,
\qquad
r = 16\epsilon_V .
\]

\item A solution is retained if it is real-valued on the chosen branch, regular in the relevant field range, and satisfies the observational constraints used in this paper. 
In the numerical tables, we use the constraints
\[
0.9607 < n_s < 0.9691,
\qquad
r < 0.056 .
\]

\item Because the Virasoro transformation involves fractional powers, different branches or roots can yield more than one valid solution for the same value of \(\theta\). 
When this happens, each valid solution is counted as an independent entry in the statistical sample.

\item Let the accepted solutions for a given pair $(n,m)$ be denoted by
\[
\{(n_{s,i},r_i)\}_{i=1}^{N_{\rm sol}} .
\]
We compute the mean values as
\[
\bar n_s = \frac{1}{N_{\rm sol}}\sum_{i=1}^{N_{\rm sol}} n_{s,i},
\qquad
\bar r = \frac{1}{N_{\rm sol}}\sum_{i=1}^{N_{\rm sol}} r_i .
\]
The standard deviations quoted in Tables~1--3 are then computed as
\[
\sigma_{n_s}
=
\sqrt{
\frac{1}{N_{\rm sol}}
\sum_{i=1}^{N_{\rm sol}}
(n_{s,i}-\bar n_s)^2
},
\qquad
\sigma_r
=
\sqrt{
\frac{1}{N_{\rm sol}}
\sum_{i=1}^{N_{\rm sol}}
(r_i-\bar r)^2
}.
\]
These quantities characterize the spread of model predictions under the scan over \(\theta\) and branch choices. 
They should not be interpreted as observational uncertainties.
\end{enumerate}
\color{black}

\section{Summary of Perturbative Calculations}

This appendix summarizes the details of the perturbative calculations
in the Virasoro-squeezing inflationary model.

\subsection{Classification of Potential Structures and Variable Transformations
by $n$}

We consider a class of inflationary models in which a scalar field
$\phi$ is subject to a Virasoro squeezing transformation defined
by

\begin{equation}
F_{n}(\phi)=\left(\frac{1}{1-n\theta\phi^{n}}\right)^{\frac{1}{n}}\phi.
\end{equation}

Here, the parameter $n$ specifies which Virasoro generator $L_{n}$
is used in the squeezing, and $\theta$ controls the strength of the
squeezing. The potential is then given by

\begin{equation}
V_{m}=\frac{1}{m}\lambda_{m}|F_{n}(\phi)|^{m}.
\end{equation}

In accordance with this transformation, we introduce a new scalar
field $\chi_{n}$ that preserves the canonical kinetic term. It is
defined through the differential relation:

\begin{equation}
\frac{d\chi_{n}}{d\phi}=\frac{1}{F'(\phi)}\Rightarrow\chi_{n}=\int^{\phi}\frac{1}{F'_{n}(\phi)}d\phi=\int^{\phi}\left(1-n\theta\phi^{n}\right)^{1+\frac{1}{n}}d\phi.
\end{equation}

This variable transformation rewrites the potential as a function
of $\chi_{n}$, which will henceforth serve as the inflaton field
in the analysis.

The behavior of this integral varies depending on the sign and magnitude
of the exponent $n$, leading to qualitatively different approximations
in different regimes. In the remainder of this section, we classify
the potential structures according to the domain of $n$, and derive
approximate expressions accordingly.

\subsubsection{Case $n>0$: Dominance in the Large-$\phi$ Regime}

In this case, the integral is dominated by the region where $\phi\to\infty$.
Therefore, we have
\begin{equation}
\chi_{n}\approx\int^{\phi}\left(-n\theta\phi^{n}\right)^{1+\frac{1}{n}}d\phi\propto\phi^{n+2}
\end{equation}
which implies the inverse relation
\begin{equation}
\phi(\chi_{n})\approx A_{n}\chi^{\frac{1}{n+2}},\ A_{n}=\left(\frac{n+2}{(-n\theta)^{1+\frac{1}{n}}}\right)^{\frac{1}{n+2}}.
\end{equation}

Accordingly, the potential can be approximated as

\begin{equation}
V_{m}\approx\frac{\lambda_{m}}{m}\left|\frac{A_{n}\chi_{n}^{\frac{1}{n+2}}}{\left(1-n\theta A_{n}^{n}\chi_{n}^{\frac{n}{n+2}}\right)^{\frac{1}{n}}}\right|^{m}.
\end{equation}

In the limit $\chi_{n}\to\infty$, this potential asymptotically approaches
a plateau, thereby naturally satisfying the slow-roll conditions required
for inflation.

\subsubsection{Case $n<-2$: Dominance in the Small-$\phi$ Regime}

When $n<0$, the integral is dominated by the region near $\phi\to0$.
In particular, for $n<-2$, we find
\begin{equation}
\chi_{n}(\phi)\approx\int^{\phi}\left(|n|\theta\phi^{-|n|}\right)^{1-\frac{1}{|n|}}d\phi\propto\phi^{2-|n|},
\end{equation}
which implies the inverse relation
\begin{equation}
\phi(\chi_{n})\approx B_{n}\chi^{\frac{1}{2-|n|}},\ B_{n}=\left(\frac{-|n|+2}{(|n|\theta)^{1-\frac{1}{|n|}}}\right)^{\frac{1}{2-|n|}}.
\end{equation}

In this case as well, the potential exhibits a plateau as $\chi_{n}\to\infty$.
Formally, the structure is nearly identical to the case $n>0$ discussed
in the previous subsection, and the derivation of inflationary slow-roll parameter
proceeds in a parallel manner.

\subsubsection{Case $-2<n<0$: Necessity of Corrections in the Intermediate Regime}

In this case, since the exponent $\frac{1}{2-|n|}>1$, the approximate
expansion $\phi(\chi_{n})$ remains valid. However, higher-order corrections
cannot be neglected when deriving the observables.

Indeed, for the integral expression of the e-folding number,
\begin{equation}
N_{e}\simeq\frac{1}{M_{pl}^{2}}\int_{\chi_{end}}^{\chi}\left(\frac{d\log V}{d\chi_{n}}\right)^{-1}d\chi_{n},
\end{equation}
a simple power-law approximation is insufficient. It becomes necessary
to incorporate correction terms such as $\chi_{n}^{n/(n+2)}$, as
well as perturbative terms like $\delta$ (discussed later).

Physically, this parameter range is also special, since the plateau
of the potential does not necessarily appear in the limit $\chi_{n}\to\infty$.
As a result, the slow-roll condition is not automatically satisfied
and must be treated with particular care.

\subsection{Observable Quantities and Slow-Roll Parameters}

The key cosmological observables, namely the spectral index $n_{s}$
and the tensor-to-scalar ratio $r$, are given under the slow-roll
approximation by the following expressions:

\begin{equation}
n_{s}-1\approx-6\epsilon_{V}+2\eta_{V},
\end{equation}

\begin{equation}
r\approx16\epsilon_{V}.
\end{equation}
Here, $\epsilon_{V}$ and $\eta_{V}$ are the slow-roll parameters,
defined with respect to the potential $V(\chi_{n})$ as

\begin{equation}
\epsilon_{V}\equiv\frac{M_{pl}^{2}}{2}\left(\frac{V_{,\chi}}{V}\right)^{2}=\frac{M_{pl}^{2}}{2}\left(\frac{\partial}{\partial\chi}\log V\right)^{2},
\end{equation}

\begin{equation}
\eta_{V}\equiv M_{pl}^{2}\frac{V_{,\chi\chi}}{V}.
\end{equation}

The e-folding number $N_{e}$, which quantifies the amount of inflation,
is defined with respect to the evolution of the inflaton field $\chi_{n}$
by the integral
\begin{equation}
N_{e}\simeq\frac{1}{M_{pl}^{2}}\int_{\chi_{end}}^{\chi}\left(\frac{d\log V}{d\chi_{n}}\right)^{-1}d\chi_{n},
\end{equation}
where $M_{pl}$ is the reduced Planck mass. The value of the field
at the end of inflation, $\chi_{end}$, is determined by the condition
\begin{equation}
\epsilon_{V}(\chi_{end})\approx1.
\end{equation}

Using these relations, the next subsection will carry out explicit
calculations of $\epsilon_{V}$ and $\eta_{V}$, and derive theoretical
predictions for the observables $n_{s}$ and $r$.

\subsection{Derivation of Slow-Roll Parameters and Analysis of Approximate Structures}

In this subsection, we explicitly derive the slow-roll parameters
$\epsilon_{V}$ and $\eta_{V}$ for the inflaton potential. We begin
by employing a simple power-law approximation for $\phi(\chi_{n})$
in order to compute the slow-roll parameters.

As discussed in Section A.1, this approximation is valid for $n>0$,
or formally applicable to the entire range $n<0$ as well. However,
in the intermediate regime $-2<n<0$, it becomes necessary to incorporate
correction terms, as will be discussed later.

\subsubsection{Evaluation of $\epsilon_{V}$}

Based on the power-law approximation $\phi(\chi_{n})\approx A_{n}\chi_{n}^{1/(n+2)}$,
the potential can be written as 

\begin{equation}
V_{m}\approx\frac{\lambda_{m}}{m}\left|\frac{A_{n}\chi_{n}^{\frac{1}{n+2}}}{\left(1-n\theta A_{n}^{n}\chi_{n}^{\frac{n}{n+2}}\right)^{\frac{1}{n}}}\right|^{m}.
\end{equation}

Taking the logarithm, we can expand $\log V_{m}$ as
\begin{equation}
\log V_{m}=\log\lambda_{m}-\log m+m\log|A_{n}|+\frac{m}{n+2}\log\left|\chi_{n}\right|-\frac{m}{n}\log|1-n\theta A_{n}^{n}\chi_{n}^{\frac{n}{n+2}}|.
\end{equation}

By differentiating this expression, the slow-roll parameter $\epsilon_{V}$
is given by
\begin{equation}
\epsilon_{V}=\frac{M_{pl}^{2}}{2}\left(\frac{\partial}{\partial\chi}\log V\right)^{2}.
\end{equation}

Explicitly differentiating $\log V_{m}$, we obtain
\begin{equation}
\frac{d\log V_{m}}{d\chi_{n}}=\frac{m}{n+2}\frac{1}{\chi_{n}}-\frac{m}{n}\left(\frac{-n\theta A_{n}^{n}\frac{n}{n+2}\chi_{n}^{\frac{-2}{n+2}}}{1-n\theta A_{n}^{n}\chi^{\frac{n}{n+2}}}\right).
\end{equation}

Therefore, the slow-roll parameter $\epsilon_{V}$ becomes
\begin{equation}
\epsilon_{V}=\frac{M_{pl}^{2}}{2}\left(\frac{m}{n+2}\frac{1}{\chi_{n}}-\frac{m}{n}\left(\frac{-n\theta A_{n}^{n}\frac{n}{n+2}\chi_{n}^{\frac{-2}{n+2}}}{1-n\theta A_{n}^{n}\chi^{\frac{n}{n+2}}}\right)\right)^{2}.
\end{equation}

This expression becomes particularly useful in the large-$\chi_{n}$
limit (i.e., during the early stage of inflation), where the dominant
terms are clearly identified and can be used for further simplification. 

\subsubsection{Evaluation of $\eta_{V}$}

The slow-roll parameter $\eta_{V}$ involves the second derivative
of the potential and is defined by

\begin{equation}
\begin{aligned}\eta_{V}=M_{pl}^{2} & \left(-\frac{m}{n+2}\frac{1}{\chi_{n}^{2}}\right.\\
 & -\frac{m}{n}\left(\frac{-n\theta A_{n}^{n}\frac{-2n}{\left(n+2\right)^{2}}\chi_{n}^{\frac{-2}{n+2}-1}}{1-n\theta A_{n}^{n}\chi^{\frac{n}{n+2}}}+\frac{-\frac{n^{3}}{(n+2)^{2}}\theta^{2}A_{n}^{2n}\chi_{n}^{\frac{-4}{n+2}}}{\left(1-n\theta A_{n}^{n}\chi^{\frac{n}{n+2}}\right)^{2}}\right)\\
 & \left.+\left(\frac{m}{n+2}\frac{1}{\chi_{n}}-\frac{m}{n}\left(\frac{-n\theta A_{n}^{n}\frac{n}{n+2}\chi_{n}^{\frac{-2}{n+2}}}{1-n\theta A_{n}^{n}\chi^{\frac{n}{n+2}}}\right)\right)^{2}\right).
\end{aligned}
\end{equation}

This expression, together with that of $\epsilon_{V}$, will be used
in the subsequent evaluation of the inflationary observables $n_{s}$
and $r$.

\subsubsection{Connection to Observables}

Using the slow-roll parameters $\epsilon_{V}$ and $\eta_{V}$, the
inflationary observables---namely, the spectral index $n_{s}$ and
the tensor-to-scalar ratio $r$---are given by

\begin{equation}
n_{s}-1\approx-6\epsilon_{V}+2\eta_{V},
\end{equation}

\begin{equation}
r\approx16\epsilon_{V}.
\end{equation}

By substituting the expressions for $\epsilon_{V}$ and $\eta_{V}$
derived in the previous subsections, these equations provide theoretical
predictions for $n_{s}$ and $r$ as functions of the inflaton field
$\chi_{n}$.

In the next subsection, we use these results to evaluate the relationship
between $\chi_{n}$ and the e-folding number, and to obtain concrete
predictions.

\subsection{Evaluation of the E-Folding Number and the Field Values $\chi,\ \chi_{end}$}

During inflation, the e-folding number $N_{e}$ is given under the
slow-roll approximation by

\begin{equation}
N_{e}\simeq\frac{1}{M_{pl}^{2}}\int_{\chi_{end}}^{\chi}\left(\frac{d\log V}{d\chi_{n}}\right)^{-1}d\chi_{n}.
\end{equation}

Here, the field value $\chi_{end}$ at the end of inflation is determined
by the condition that the slow-roll parameter $\epsilon_{V}$ reaches
unity: 

\begin{equation}
\epsilon_{V}(\chi_{end})\approx1.
\end{equation}

Using the approximate expression of $\epsilon_{V}(\chi_{n})$ derived
earlier, this condition can be treated as an equation for $\chi_{end}$,
which can then be solved approximately.

Throughout the following discussion, we introduce the notation $(\pm)_{(0)}$
to explicitly indicate the sign of $\chi_{end}$. Although the actual
sign may vary depending on the context, we adopt this notation for
notational consistency, under the assumption that the physical field
magnitude is treated as positive.

\subsubsection{Case $n>0$ and $n<-2$}

In this case, the field value $\chi_{n}$ is sufficiently large, and
the condition $n\theta A_{n}^{n}\chi_{n}^{n/(n+2)}\ll1$ holds. Under
this assumption, the following approximation is valid:

\begin{equation}
\frac{d\log V_{m}}{d\chi_{n}}\approx\frac{m}{n+2}\frac{1}{\chi_{n}}-\frac{m}{n}\left(\frac{-n\theta A_{n}^{n}\frac{n}{n+2}\chi_{n}^{\frac{-2}{n+2}}}{1-n\theta A_{n}^{n}\chi^{\frac{n}{n+2}}}\right).
\end{equation}

The e-folding number can then be evaluated as
\begin{equation}
N_{e}\simeq\frac{1}{M_{pl}^{2}}\int_{\chi}^{\chi_{end}}\left(\frac{m}{n+2}\frac{1}{\chi_{n}}-\frac{m}{n}\left(\frac{-n\theta A_{n}^{n}\frac{n}{n+2}\chi_{n}^{\frac{-2}{n+2}}}{1-n\theta A_{n}^{n}\chi^{\frac{n}{n+2}}}\right)\right)^{-1}d\chi_{n}.
\end{equation}

This integral can be approximated in the limit $\chi_{n}\to\infty$,
and it reduces to a power-law integral based on the variable transformation
$\phi\sim\chi_{n}^{1/(n+2)}$. As a result of this approximation,
we obtain the following relation:
\begin{equation}
N_{e}\simeq\frac{1}{M_{pl}^{2}}\frac{2n(n+1)}{m}\theta A_{n}^{n}\left(\chi_{end}^{\frac{n}{n+2}}-\chi^{\frac{n}{n+2}}\right).
\end{equation}

This expression represents the difference between $\chi_{end}$ and
$\chi$ with an exponent depending on $n$, and can be used to determine
the inflaton field value at the beginning of inflation.

Furthermore, from the condition $\epsilon_{V}(\chi_{end})\approx1$,
we obtain the following approximate equation:
\begin{equation}
\frac{m}{n+2}\frac{1}{\chi_{end}^ {}}-\frac{m}{n}\left(\frac{-n\theta A_{n}^{n}\frac{n}{n+2}\chi_{end}^{\frac{-2}{n+2}}}{1-n\theta A_{n}^{n}\chi_{end}^{\frac{n}{n+2}}}\right)\approx(\pm)_{(0)}\frac{\sqrt{2}}{M_{pl}}.
\end{equation}

Assuming that $\chi_{end}$ is sufficiently large such that $n\theta A_{n}^{n}\chi_{end}^{n/(n+2)}\ll1$,
we can expand the denominator to leading order and approximate:
\begin{equation}
-\frac{m}{n+2}\frac{1}{n\theta}A_{n}^{-n}\chi_{end}^{-1}\approx(\pm)_{(0)}\frac{\sqrt{2}}{M_{pl}}.
\end{equation}
Solving for $\chi_{end}$, we obtain:
\begin{equation}
\chi_{end}\approx-(\pm)_{(0)}\frac{n(n+2)}{m}\frac{M_{pl}}{\sqrt{2}}n\theta A_{n}^{-n}.
\end{equation}

Substituting the above expression for $\chi_{end}$ into the previously
derived formula for the e-folding number, we obtain the following
expression for $\chi$:

\begin{equation}
\chi=\chi_{end}\left(1-\frac{2n(n+1)}{mN_{e}M_{pl}^{2}}\theta A_{n}^{n}\chi_{end}^{\frac{n}{n+2}}\right).
\end{equation}

This leads to a structure with a power-law correction:
\begin{equation}
\chi=-(\pm)_{(0)}\frac{n(n+2)}{m}\frac{M_{pl}}{\sqrt{2}}n\theta A_{n}^{-n}\left(1-\frac{2n(n+1)}{mN_{e}M_{pl}^{2}}\theta A_{n}^{n}\left(-(\pm)_{(0)}\frac{n(n+2)}{m}\frac{M_{pl}}{\sqrt{2}}n\theta A_{n}^{-n}\right)^{\frac{n}{n+2}}\right).
\end{equation}

This expression serves as a fundamental formula for estimating the
value of the inflaton field $\chi$ at the beginning of inflation
in terms of the e-folding number $N_{e}$.

\subsubsection{Case $-2<n<0$}

In this case, the field value $\chi_{n}$ is not necessarily large,
and the quantity $n\theta A_{n}^{n}\chi_{n}^{n/(n+2)}$ is not guaranteed
to be small. As a result, the approximations used in the previous
subsection---where leading-order terms cancel and correction terms
dominate---are no longer applicable, and a different treatment is
required.

Under this condition, the e-folding number can be approximated as
\begin{equation}
N_{e}\simeq\frac{1}{M_{pl}^{2}}\frac{m}{n+2}\left(\frac{1}{2}\chi_{end}^{2}-\frac{1}{2}\chi^{2}-n\left(\frac{n}{n+2}+1\right)\theta A_{n}^{n}\left(\chi_{end}^{\frac{n}{n+2}+2}-\chi_{n}^{\frac{n}{n+2}+2}\right)\right).
\end{equation}

This expression arises from an approximate integration with respect
to $\chi_{n}$ \LyXZeroWidthSpace , in which the leading-order power-law
term $\chi^{2}$ and the correction term $\chi^{n/(n+2)}$ are treated
separately.

By inverting the above expression, we obtain an approximate expansion
of $\chi$ as a function of the e-folding number $N_{e}$:
\begin{equation}
\chi\approx\chi_{end}\left(1-\frac{2(n+2)}{m}N_{e}M_{pl}^{2}\chi_{end}^{-2}-2n\frac{2n+2}{n+2}\theta A_{n}^{n}\chi_{end}^{\frac{n}{n+2}}\left(\chi_{end}^{\frac{n}{n+2}}-\chi^{\frac{n}{n+2}}\right)\right)^{\frac{1}{2}}.
\end{equation}

Expanding the power and reorganizing the terms perturbatively, we
arrive at a simpler form:

\begin{equation}
\chi\approx\chi_{end}\left(1-\frac{(n+2)}{m}N_{e}M_{pl}^{2}\chi_{end}^{-2}-n\frac{2n+2}{n+2}\theta A_{n}^{n}\chi_{end}^{\frac{n}{n+2}}\left(\chi_{end}^{\frac{n}{n+2}}-\chi^{\frac{n}{n+2}}\right)\right).
\end{equation}

Moreover, the value of $\chi_{end}$ is approximately determined by
the condition $\epsilon_{V}(\chi_{end})\approx1$, leading to
\begin{equation}
\chi_{end}\approx\left(\pm\right)_{0}\frac{n+2}{m}\frac{M_{pl}}{\sqrt{2}}\left(1+\left(\left(\pm\right)_{0}\right)^{\frac{n}{n+2}}n\theta A_{n}^{n}\left(\frac{n+2}{m}\frac{M_{pl}}{\sqrt{2}}\right)^{\frac{n}{n+2}}\right).
\end{equation}

Accordingly, $\chi$ can also be expanded in the same manner as
\begin{equation}
\chi\approx\left(\pm\right)_{0}\frac{n+2}{m}\frac{M_{pl}}{\sqrt{2}}\left(1+\left(\left(\pm\right)_{0}\right)^{\frac{n}{n+2}}n\theta A_{n}^{n}\left(\frac{n+2}{m}\frac{M_{pl}}{\sqrt{2}}\right)^{\frac{n}{n+2}}\right)\left(1-2\frac{mN_{e}}{n+2}\right).
\end{equation}

In this way, for the case $-2<n<0$, it is necessary to explicitly
retain both the power-law correction and the perturbative contributions.
This requires a different treatment from that in the previous subsection.

\subsection{Crude Estimation}

In this subsection, we use the approximate expressions for $\chi$
and $\chi_{end}$ derived in the previous subsection, along with the
leading-order contributions of the slow-roll parameters, to obtain
a crude estimate of the inflationary observables $n_{s}$ and $r$.

These observables are given in terms of the slow-roll parameters $\epsilon_{V}$
and $\eta_{V}$ as follows:

\begin{equation}
n_{s}\approx1-6\epsilon_{V}+2\eta_{V},\ r\approx16\epsilon_{V}.
\end{equation}

Under the power-law approximation, we may assume that $\chi$ is sufficiently
large, which allows the following leading-order approximations for
the slow-roll parameters:
\begin{equation}
\epsilon_{V}\approx\frac{M_{pl}^{2}}{2}\left(\frac{m}{n+2}\frac{1}{\chi_{}}-\frac{m}{n}\left(\frac{\boldsymbol{-}\frac{n}{n+2}\chi_{}^{-1}}{n^{-1}\theta^{-1}A_{n}^{-n}\chi^{-\frac{n}{n+2}}-1}\right)\right)^{2},
\end{equation}

\begin{equation}
\begin{aligned}\eta_{V}\approx & M_{pl}^{2}\left(-\frac{m}{n+2}\frac{1}{\chi^{2}}\right.\\
 & -\frac{m}{n}\left(\frac{-n\theta A_{n}^{n}\frac{-2n}{\left(n+2\right)^{2}}\chi_{}^{\frac{-2}{n+2}-1}}{1-n\theta A_{n}^{n}\chi^{\frac{n}{n+2}}}+\frac{-\frac{n^{3}}{(n+2)^{2}}\theta^{2}A_{n}^{2n}\chi_{n}^{\frac{-4}{n+2}}}{\left(1-n\theta A_{n}^{n}\chi^{\frac{n}{n+2}}\right)^{2}}\right)\\
 & \left.+\left(\frac{m}{n+2}\frac{1}{\chi}-\frac{m}{n}\left(\frac{-n\theta A_{n}^{n}\frac{n}{n+2}\chi^{\frac{-2}{n+2}}}{1-n\theta A_{n}^{n}\chi^{\frac{n}{n+2}}}\right)\right)^{2}\right).
\end{aligned}
\end{equation}

The field value $\chi$ in these expressions can be substituted using
the approximate solution obtained in the previous subsection. The
observables $n_{s}$ and $r$ can then be evaluated either numerically
or through these analytical approximations.

\subsection{Estimation of Errors}

We now estimate the accuracy of the approximation
\begin{equation}
\phi(\chi_{n})\approx A_{n}\chi_{n}^{1/(n+2)}
\end{equation}
which has been used throughout the previous analysis. To do so, we
consider a perturbed form with a correction term $\delta$:
\begin{equation}
\phi(\chi_{n})\approx(A_{n}\chi_{n}^{\frac{1}{n+2}})\left(1+\delta\right),
\end{equation}
and evaluate perturbatively how accurate this approximation is.

Consider the integral expression:
\begin{equation}
\chi_{n}=\int^{\phi}\left(1-n\theta\phi^{n}\right)^{1+\frac{1}{n}}d\phi.
\end{equation}

By expanding the structure of the correction term in this integral,
the perturbation $\delta$ is given by
\begin{equation}
\delta=\frac{n+1}{2n^{2}\theta}\left(A_{n}^{-n}\chi^{-\frac{n}{n+2}}\right).
\end{equation}

In this case, the potential can be written as
\begin{equation}
V(\chi_{n})=V_{0}(\chi_{n})\tilde{\delta},
\end{equation}

where 
\begin{equation}
\tilde{\delta}=1+K\delta
\end{equation}

with
\begin{equation}
K=\frac{m}{1-n\theta A_{n}^{n}\chi_{n}^{n/(n+2)}}.
\end{equation}

This leads to the following decomposition for the logarithm of the
potential:

\begin{equation}
\log V=\log V_{0}+\log\tilde{\delta}.
\end{equation}

The resulting corrections to the slow-roll parameters can then be
evaluated as
\begin{equation}
\Delta\epsilon_{V}=M_{pl}^{2}\left(\frac{\partial}{\partial\chi}\log V_{0}\right)\left(\frac{\partial}{\partial\chi_{n}}\ln\tilde{\delta}\right),
\end{equation}
\begin{equation}
\Delta\eta_{V}=2M_{pl}^{2}\left(\frac{\partial}{\partial\chi}\log V_{0}\right)\left(\frac{\partial}{\partial\chi_{n}}\ln\tilde{\delta}\right)+M_{pl}^{2}\frac{\partial^{2}}{\partial\chi_{n}^{2}}\ln\tilde{\delta}.
\end{equation}

Using these expressions, the corrections to the observable quantities
can be estimated as
\begin{equation}
\Delta n_{s}\approx-6\Delta\epsilon_{V}+2\Delta\eta_{V},
\end{equation}

\begin{equation}
\Delta r\approx16\Delta\epsilon_{V}.
\end{equation}

When $\delta\ll1$, these corrections provide a sufficiently accurate
approximation while neglecting higher-order terms. Therefore, $\delta$
serves as one of the primary indicators of theoretical uncertainty
in the model predictions.

\section{Validity Range of the Squeezing Parameter $\theta$}

We estimate the valid range of the Virasoro--squeezing parameter $\theta$ and discuss its relation to the observed quantities $(A_s, r)$. This appendix provides a quantitative justification for the natural interval of $\theta$ constrained simultaneously by observational data and regularity (absence of branch singularities).

\subsection{Slow--roll approximation and curvature power spectrum}

Under the slow--roll approximation, the Hubble parameter $H$ and the potential $V(\phi)$ satisfy
\begin{align}
3 M_{\rm pl}^2 H^2 &\simeq V, \\
3H\dot{\phi} &\simeq -V'.
\end{align}
The slow--roll parameters are defined as
\begin{align}
\epsilon_V = \frac{M_{\rm pl}^2}{2}\left(\frac{V'}{V}\right)^{2}, \qquad 
\epsilon_H = -\frac{\dot{H}}{H^{2}} \simeq \epsilon_V .
\end{align}
The tensor--to--scalar ratio is then given by
\begin{align}
r = 16 \epsilon_H c_s \simeq 16 \epsilon_V c_s,
\end{align}
where $c_s$ denotes the sound speed of scalar perturbations.

The power spectrum of curvature perturbations is expressed as
\begin{align}
\mathcal{P}_{\mathcal{R}}(k) = 
\frac{H^{2}}{8\pi^{2} M_{\rm pl}^{2}\epsilon_H c_s}
\simeq \frac{V}{24\pi^{2} M_{\rm pl}^{4}\epsilon_V c_s}.
\end{align}
At the pivot scale $k_0 = 0.002\,{\rm Mpc}^{-1}$, we use
\begin{equation}
A_s \equiv \mathcal{P}_{\mathcal{R}}(k_0) 
\simeq 2.1\times 10^{-9}.
\end{equation}

For the tensor-to-scalar ratio, we may take
\begin{equation}
r_{\rm ref}(k_0) \lesssim 0.03
\end{equation}
as a representative reference value when estimating the inflationary energy
scale. This value is not imposed as an additional constraint in the numerical
scan performed in the main text.

\subsection{Virasoro--squeezed potential and plateau height}

The Virasoro--squeezed potential is given by
\begin{align}
V_\chi(\chi) = V(|F_n(\phi(\chi))|^2), \qquad 
F_n(\phi(\chi)) = \frac{\phi(\chi)}{(1 - n\theta\,\phi(\chi)^n)^{1/n}}.
\end{align}
From the maximum--modulus principle (no local extrema inside the regular domain) and the asymptotic behavior of $F_n$, one obtains
\begin{align}
|F_n(\phi(\chi))| \xrightarrow[]{|\phi|\to\infty} |n\theta|^{-1/n}.
\end{align}
Thus, an $m$-th order potential takes the plateau form during slow roll:
\begin{align}
V(\chi) = \frac{\lambda_m}{m}\,|F_n(\phi(\chi))|^{m}
\;\approx\; \frac{\lambda_m}{m}\,|n\theta|^{-m/n}.
\end{align}

\subsection{Relation to the observed quantities}

Using the expression for $A_s$,
\begin{align}
A_s = \frac{V}{24\pi^{2} M_{\rm pl}^{4}\epsilon_V c_s}
\simeq \frac{V}{24\pi^{2} M_{\rm pl}^{4}r c_s/16}
= \frac{2}{3}\frac{\pi^2 V}{M_{\rm pl}^{4}c_s}\frac{1}{r}.
\end{align}
At the pivot scale, the corresponding plateau potential is
\begin{align}
V_{\rm pl} = \frac{3}{2}\pi^{2}A_s r M_{\rm pl}^{4}c_s.
\end{align}
Equating this with the plateau height of the Virasoro--squeezed potential gives
\begin{align}
\frac{\lambda_m}{m}\,|n\theta|^{-m/n}
= \frac{3}{2}\pi^{2}A_s r M_{\rm pl}^{4}c_s,
\end{align}
which yields
\begin{align}
|n\theta|^{-\frac{m}{n}}
= \frac{3}{2}\pi^{2}\frac{m}{\lambda_m}A_s r M_{\rm pl}^{4}c_s.
\end{align}
Hence,
\begin{align}
\boxed{
|\theta|
= \frac{1}{|n|}
\left(\frac{\lambda_m}{m}\right)^{\frac{n}{m}}
\left(\frac{3}{2}\pi^{2}A_s r M_{\rm pl}^{4}c_s\right)^{-\frac{n}{m}}
}.
\end{align}

\subsection{Scaling relations}

From the above expression, the scaling of the squeezing parameter is
\begin{align}
|\theta| \propto r^{-\frac{n}{m}}, 
\qquad 
|\theta| \propto A_s^{-\frac{n}{m}}.
\end{align}
For example, for $m=2$:
\begin{align}
n=-2:\; |\theta| \propto r^{+1},\,A_s^{+1},
\qquad
n>0:\; |\theta| \propto r^{-n/2}.
\end{align}
Thus, for $n=-2$, a larger $r$ leads to a larger $|\theta|$, whereas for $n>0$, smaller $r$ requires larger $|\theta|$.

\subsection{Regularity constraint and sign of $\theta$}

Since $F_n$ has a branch point at
\begin{align}
\phi_c = |(n\theta)^{-1/n}|,
\end{align}
the field value at the end of inflation $\phi_{\rm end}$ (defined by $\epsilon_V(\phi_{\rm end})=1$) must lie inside the regular domain:
\begin{align}
|\phi_{\rm end}| < |\phi_c|.
\end{align}
This gives the upper bound
\begin{align}
\boxed{
|\theta| < \frac{|\phi_{\rm end}|^{-n}}{|n|}.
}
\end{align}
The sign of $\theta$ is fixed by regularity along the real axis:
\begin{align}
\boxed{
n>0 \Rightarrow \theta<0, \qquad
n<0 \Rightarrow \theta>0.
}
\end{align}

\subsection{Summary}

The squeezing parameter $\theta$ is therefore constrained by two independent conditions:

\begin{enumerate}
\item \textbf{Observational constraint}
\begin{align}
|\theta| =
\frac{1}{|n|}
\left(\frac{\lambda_m}{m}\right)^{\frac{n}{m}}
\left(\frac{3}{2}\pi^{2}A_s r M_{\rm pl}^{4}c_s\right)^{-\frac{n}{m}}.
\end{align}

\item \textbf{Regularity constraint}
\begin{align}
|\theta| < \frac{|\phi_{\rm end}|^{-n}}{|n|}.
\end{align}
\end{enumerate}

When these conditions are satisfied simultaneously---together with the sign rule above---the parameter $\theta$ acquires a natural and non--arbitrary range determined entirely by the observational quantities $(A_s, r)$ and by the mathematical requirement of analyticity of the Virasoro--squeezed transformation.

\bibliographystyle{unsrt} 
\bibliography{20250803_virasoro_inflation_katagiri_v4}

\end{document}